\documentclass[twocolumn,aps,prx,superscriptaddress,nofootinbib]{revtex4-1}
\usepackage{graphicx}
\usepackage{times}
\usepackage{bm}
\usepackage[linktocpage=true]{hyperref}
\usepackage{pgfplots}
\usepackage{amsmath}
\usetikzlibrary{arrows}
\usepackage{enumerate}
\usetikzlibrary[patterns] % ConTEXt when using Tik Z
\usetikzlibrary[snakes] % ConTEXt when using Tik Z
\usetikzlibrary[shapes] % ConTEXt when using Tik Z
\usepackage{pgfplots}

\begin{document}

\title{Proximity-induced low-energy renormalization in hybrid semiconductor-superconductor Majorana structures}
\author{Tudor D. Stanescu}
\affiliation{Department of Physics and Astronomy, West Virginia University, Morgantown, WV 26506}
\affiliation{Condensed Matter Theory Center and Joint Quantum Institute, Department of Physics, University of Maryland, College Park, Maryland, 20742-4111, USA}
\author{Sankar Das Sarma}
\affiliation{Condensed Matter Theory Center and Joint Quantum Institute, Department of Physics, University of Maryland, College Park, Maryland, 20742-4111, USA}

\begin{abstract}
A minimal model for the hybrid  superconductor-semiconductor nanowire Majorana platform is developed that  fully captures  the effects of the low-energy renormalization of the nanowire modes arising from the presence of the parent superconductor.  In this model, the parent superconductor is an active component that participates explicitly in the low-energy physics, not just a passive partner that only provides proximity-induced Cooper pairs for the nanowire. This treatment on an equal footing of the superconductor and the semiconductor  has become necessary in view of recent experiments, which do not allow a consistent interpretation based just on the bare semiconductor properties.  The general theory involves the evaluation of the exact semiconductor Green function that includes a dynamical self-energy correction arising from the tunnel-coupled superconductor. Using a tight binding description, the nanowire Green function is obtained in various relevant parameter regimes,  with the parent superconductor being treated within the BCS-BdG prescription.  General conditions for the emergence of topological superconductivity are worked out for single-band as well as multi-band nanowires and detailed numerical results are given for both infinite and finite wire cases.  The topological quantum phase diagrams are provided numerically and the Majorana bound states are obtained along with their oscillatory energy splitting behaviors due to wavefunction overlap in finite wires.  Renormalization effects  are shown to be both qualitatively and quantitatively important in modifying the low-energy spectrum of the nanowire.  The results of the theory are found to be in good qualitative agreement with  Majorana nanowire experiments, leading to the conclusion that the proximity-induced low-energy renormalization of the nanowire modes by the parent superconductor is of fundamental importance in superconductor-semiconductor hybrid structures, except perhaps in the uninteresting limit of extremely weak superconductor-semiconductor  tunnel coupling. Implications of the general theory for obtaining true zero energy topological Majorana modes are pointed out.
\end{abstract}

\maketitle

\section{Introduction}

Study of Majorana fermions or, more precisely,  Majorana zero energy modes (often also called 'Majorana bound states'), along with the closely related topic of topological superconductivity is among the most active current research areas in physics, attracting interest not only from theoretical and experimental physicists, but also from materials scientists, computer scientists and engineers, mathematicians, and electrical engineers \cite{Nayak2008, Stanescu2017}.  Majorana modes are the robust  gapless (i.e. zero-energy) boundary states that form naturally at the surfaces (or the edges) of bulk-gapped topological superconductors.The broad inter-disciplinary interest in the subject arises from the possibility of  using a Majorana platform to carry out topological quantum computation, which is fault-tolerant and, in principle, does not necessitate quantum error correction \cite{Kitaev2001,DSarma2015}. More importantly, there is a deep aspect of the physics at play here: the fact that localized Majorana zero modes obey non-Abelian anyonic (i.e. neither fermionic nor bosonic \cite{Leinaas1977,Wilczek1982}) braiding statistics and are therefore strange topological objects of fundamental interest in their own right, quite distinct from their proposed applications in topological quantum computation.  

Although there are many proposed techniques for creating Majorana modes in the laboratory \cite{Alicea2012,Leijnse2012,Beenakker2013,Stanescu2013}, 
such as fractional quantum Hall effects \cite{Read2000,Nayak1996,DSarma2005} and topological insulators in contact with superconductors \cite{Fu2008},  the method that has attracted the most experimental interest involves a semiconductor-superconductor hybrid structure \cite{Sau2010a}, where, through proximity effect,  a parent superconductor induces topological superconductivity in a strongly spin-orbit coupled semiconductor in the presence of an applied magnetic field.  The particular system that is actively studied in many laboratories \cite{Mourik2012,Rokhinson2012,Deng2012,Das2012,Churchill2013,Finck2013,Chang2014,Krogstrup2015,Albrecht2016,Deng2016,Zhang2016,Chen2016}
 is a one-dimensional semiconductor nanowire (e.g., InSb, InAs) with strong spin-orbit coupling in proximity to a bulk superconductor (e.g., Al, NbTiN) and  in the presence of an external magnetic field applied along the wire, or, more generally,  along a direction perpendicular to the spin-orbit coupling field.  It has been established theoretically that such a system becomes a topological superconductor when the Zeeman spin splitting in the nanowire is above a certain critical value \cite{Sau2010a,Sau2010,Lutchyn2010,Oreg2010}.
 In fact,   at this critical value of the spin splitting the system  undergoes a topological quantum phase transition \cite{Read2000,Sau2010a}
 from an ordinary, topologically-trivial superconductor (with a bulk gap, but no boundary states) to a topological superconductor with a bulk gap and Majorana modes at the wire ends.  
In long wires, these Majorana modes manifest themselves as robust midgap zero-energy states inside the bulk topological superconducting gap. 
For short wires, the end Majorana modes may couple to each other producing oscillatory energy splittings away from zero as a function of wire length, chemical potential,  and/or magnetic field.
  The topological gap itself turns out to be proportional to the strength of the effective spin-orbit coupling, while the critical field strength (or more precisely, the critical spin-splitting) depends on the effective coupling strength between the semiconductor and the superconductor and on the chemical potential of the wire.  Theoretical research shows that this phenomenon of proximity-induced topological superconductivity (and the associated creation of Majorana zero modes), which requires the presence of proximity effect, Zeeman spin splitting, and spin-orbit coupling, is stable to realistic effects of disorder \cite{Sau2012,Weperen2015},  interaction \cite{Lobos2012,Stoudenmire2011,Thomale2013}, and multichannel or multisubband effects in the nanowire (i.e. the nanowire not being in the strict one-dimensional limit) \cite{Lutchyn2011,Stanescu2011}, provided that these effects are not too strong.  There are now many concrete proposals to use these nanowires as parts of elaborate topological quantum circuits in order to carry out Majorana braiding, fusion, and measurement so that practical topological quantum computing becomes feasible in the near future \cite{Bonderson2009,Alicea2011,Hyart2013,Aasen2016,Bonderson2016,Hastings2016,Vijav2016,Landau2016,Karzig2016,Plugge2017}. 

Much excitement stems from the fact that the theoretical proposals and predictions for creating topological superconductivity in semiconductor wire-superconductor hybrid structures have led to experiments showing evidence for the possible existence of Majorana zero modes at the wire ends \cite{Mourik2012,Deng2012,Das2012,Churchill2013,Finck2013,Chang2014,Deng2016,Zhang2016,Chen2016}.  A serious problem, however, has been the fact that, in spite of six years of intense experimental activity that has uncovered some signs for the possible existence of Majorana modes, no definitive conclusion  can be drawn. In addition, there are many nagging discrepancies between theory and experiment casting a dark cloud over the whole subject.  One example of such a discrepancy is the persistence of a soft gap \cite{Stanescu2014}  with considerable subgap features, which greatly complicate the Majorana interpretation of the results, since the Majorana mode is itself a particular subgap state (albeit, in ideal conditions, precisely at midgap).  Following theoretical suggestions for improving the superconductor-semiconductor interface in order to optimize the proximity effect \cite{Takei2013}, there has been substantial progress in creating 'hard' proximity gaps in nanowires \cite{Chang2014,Zhang2016}
 at zero or low magnetic field values. However, these hard gaps disappear in the putative finite-field topological regime where the Majorana mode presumably appears.  In fact, all experiments reporting evidence for Majorana zero modes suffer from the gap being extremely soft precisely in the regime where the evidence for the Majorana zero modes emerges, notwithstanding the achievement of a hard gap in the low-field, topologically-trivial regime.  This unwanted feature occurs in spite of the nanowires being relatively disorder-free and essentially ballistic, as far as their zero-field transport properties are concerned.  

A closely related problem is that the experimental Majorana zero mode, which is observed  as a zero-bias conductance peak in a charge tunneling measurement, seems to have a considerable width around zero energy.  This broadening is comparable to the (soft) topological gap itself \cite{Deng2016,Zhang2016,Chen2016}!  This is in sharp contrast to the theoretical prediction of a sharp zero energy Majorana mode, with resolution-limited broadening in long wires, signaled by a zero-bias conductance peak that is quantized at zero temperature and has a width that is only controlled by temperature and by the transparency of the tunnel barrier.  Although there have been ad hoc attempts \cite{DSarma2016,Liu2016} to explain these disturbing features by introducing 'dissipation' as the possible broadening mechanism, the source and the microscopic physics for such Majorana-broadening (which is much larger than the thermal broadening) remains unknown.  A broad Majorana mode centered around zero energy is unsuitable for topological quantum computation (and, strictly speaking, it is not a non-Abelian anyon).  For example, this broadening may simply be hiding a large splitting of the Majorana mode into two (or more) finite-energy modes due to the wavefunction overlap between Majorana modes localized around different points along the wire. A far more dangerous possibility is that the broad peak may be associated with a cluster of low-energy-fermionic hybrid states representing combinations of the Majorana mode with sub-gap states from the parent superconductor. If so, then the broad zero-energy Majorana features observed experimentally are closely connected with the renormalization of the low-energy spectrum,  both effects being generated by the parent superconductor.
In hybrid systems where the Majorana broadening is comparable to the topological gap, which is the actual situation suggested by the experimental results reported so far, the possibility of successful anyonic braiding experiments and eventual topological quantum computation becomes hopeless \cite{Aasen2016,Clarke2016}.  

Another serious problem has been the absence in the experimental data of the predicted Majorana splitting oscillations \cite{Cheng2009,DSarma2012,Rainis2013}, which should be ubiquitously present in short-enough wires as the magnetic field is increased. 
 Actually, a recent experiment \cite{Albrecht2016} has reported the observation of Majorana oscillations (although somewhat indirectly, through Coulomb blockade peaks), but these oscillations seem to decrease with increasing magnetic field in direct conflict with the theoretical predictions.  In addition, the experiments often see many prominent extra features (e.g., gaps opening and closing as functions of both magnetic field and gate voltage), which are simply not present in the minimal theories used extensively for describing superconductor-semiconductor Majorana nanowire systems.  Independent of the important question of whether the topological Majorana zero mode has actually been observed experimentally or not, it has become clear that the existing theoretical models are unable to provide even a zeroth order systematic description for the experimental data, e.g., at the level of the qualitative understanding of the induced superconducting proximity effect itself.  The current work is devoted to the development of the appropriate zeroth order model capable of describing the low-lying energy spectrum of the nanowire in the presence of the superconducting proximity-effect induced by the parent superconductor. The main ingredient of this new theory is that it includes the parent superconductor as an {\em active  ingredient} in the physics of the heterostructure, instead of considering it simply as an inert source for producing Cooper pairs that  travel into the nanowire.  Our main message is that the parent superconductor has to be included {\em explicitly} into the theory, even at the zeroth order.  
In particular, we will bring  considerable evidence that the parent superconductor plays a key role in renormalizing all low energy properties of the nanowire. The rather ambitious goal of the current paper is to establish the minimal model for  semiconductor-superconductor hybrid structures
by providing  detailed results regarding the low-energy physics of the system that include renormalization effects arising from the presence of the parent superconductor.  We believe that, in addition to introducing the  minimal model for the study of semiconductor-superconductor hybrid structures, the current work provides strong evidence that the parent superconductor plays a crucial role in determining the low-energy physics of the hybrid system (i.e., the properties of the system at energies lower than the parent superconducting gap) and, consequently, treating it as an {\em active ingredient}  represents a {\em necessary} step toward understanding the features present in the experimental works on Majorana nanowires.

The possibility of the parent superconductor playing a nontrivial role in determining the nanowire properties has been mentioned and discussed in a few specific instances in a somewhat piecemeal manner \cite{Sau2010,Sau2012a,DSarma2015a,Cole2015,Cole2016}.  In particular, the fact that disorder in the parent superconductor could drastically (and adversely) affect the nanowire topological behavior (at least) in the strongly tunnel-coupled superconductor-semiconductor hybrid system has been already emphasized theoretically \cite{Hui2015,Cole2016}.  Our goal, however, is to develop the essential minimal model for describing the hybrid structure, where the superconductor could, for example, couple differently to different subband levels in the nanowire and, more importantly, could generate proximity-induced inter-subband coupling. This leads to qualitatively new physics, which is absent in the extensively-used minimal model  where the proximity effect is simply incorporated as an ad hoc pairing  term in the nanowire Hamiltonian.
Our approach is to treat the semiconductor, the superconductor, and the semiconductor-superconductor coupling on an equal footing right at the Hamiltonian level [see  Eq. (1) in Sec. II, for example].  The Green function description [see Eq. (3)] is equivalent to the Hamiltonian model and does not involve any additional approximation. 
We make approximations, of course, motivated by physical and numerical considerations, but these approximations, which are included in the model Hamiltonian, can be systematically relaxed if experimental results warrant such improvement, at the cost of more demanding numerical calculations.  We show that treating the parent superconductor explicitly,  at the Hamiltonian level or as a self-energy contribution to the Green function, introduces qualitatively new features in our theory, which are (1) in agreement with experiments, and (2) cannot really be incorporated  in an ad hoc manner.  For example, the near-zero energy midgap spectrum of the nanowire (i.e. the part of most interest to Majorana zero mode considerations) could be drastically different from what one finds in the minimal model where the proximity effect is simply put in by hand.  We find the renormalization effects introduced by the superconductor to be subtle and often intuitively non-obvious since the superconductor may very well affect different energy levels (and their coupling) differently in the nanowire.  Although the renormalization effects are obviously much more important in the strong-coupling regime where the coupling at the superconductor-semiconductor interface is large, we find nontrivial renormalization also in the intermediate coupling regime.  It is only in the extreme weak-coupling regime, where the induced superconducting gap is very small, that we recover the results of the standard extensively used minimal model, but this limit may not be of much practical significance since one is interested in having as large an induced bulk gap as possible because the  topological protection of the Majorana modes is supplied entirely by the size of the energy gap. Experimental systems with very small induced superconducting gaps are, therefore, of rather limited interest, while the extremely weak coupling regime, where the commonly-used minimal model strictly applies, is of purely academic interest.

The rest of the paper is organized as follows.  In section II, we introduce a minimal model for describing the hybrid structure that explicitly includes both the parent superconductor and the semiconductor nanowire, which are treated on an equal footing.  In Sec. III, we study the infinite wire case using this new model, focusing on the induced gap and the phase diagram.  These results are compared  with the finite wire situation presented in Sec. IV, which is the bulk of the paper since it corresponds to the experimentally relevant nanowires.  We conclude in section V with a summary of our main findings as well as a discussion of open questions and possible future directions.  A large number of numerical results are presented throughout the text showing the low energy spectrum of the nanowire in order to explicitly bring out the key importance of the superconductor-induced renormalization.

\section{Theoretical model}

In general, the Hamiltonian describing a semiconductor wire proximity-coupled to a conventional s-type superconductor  has four basic terms,
\begin{equation}
H = H_{\rm SM} + H_{\rm SC} + H_{\rm SM-SC} + H_{\rm ext},                            \label{Htot}
\end{equation}
where $H_{\rm SM}$ and $H_{\rm SC}$ are the Hamiltonians of the semiconductor (SM) wire and of the superconductor (SC), respectively, $H_{\rm SM-SC}$ describes the coupling between the two components of the hybrid structure, and $H_{\rm ext}$ includes contributions from external fields, such as magnetic fields and electrostatic potentials. The specific form of each term in Eq. (\ref{Htot}) depends on the details of the structure (e.g., materials, geometry, etc.) and on the degree of complexity that we want to incorporate into the model (e.g., number of bands, presence of disorder, etc.). Using tight binding models is particularly convenient for describing the SM-SC coupling. Let us assume that the SM wire is defined on a certain lattice ${\cal L}_{\rm SM}$, while the superconductor is defined on ${\cal L}_{\rm SC}$. The generic form of the coupling term is 
\begin{equation}
H_{\rm SM-SC} = \sum_{i, \alpha}\sum_{\ell, \sigma} \left(\tilde{t}_{i \ell}^{\alpha \sigma}~\! c_{i\alpha}^\dagger a_{\ell\sigma} + {\rm h.c.}\right),                                                                            \label{Hsmsc}
\end{equation}  
where $i\in {\cal L}_{\rm SM}$ and $\ell\in {\cal L}_{\rm SC}$ are lattice sites near the SM-SC interface,  $\alpha$ labels orbital and spin degrees of freedom corresponding to the SM wire, and $\sigma$ represents the orbital and/or spin degrees of freedom of the superconductor. The annihilation operators for electrons inside the SM and the SC are $c_{i\alpha}$ and $a_{\ell\sigma}$, respectively, and the hopping across the SM-SC interface is characterized by the elements $\tilde{t}_{i \ell}^{\alpha \sigma}$ of a coupling matrix $\widetilde{T}$.

The low-energy physics of the proximity-coupled SM wire can be described in terms of the Green function $G(\omega) = (\omega - H)^{-1}$ corresponding to the total Hamiltonian from Eq. (\ref{Htot}). In the real space representation $G$ is a matrix labeled by a set of parameters $m=\{r, \lambda, \tau\}$, where $r\in{\cal L}_{\rm SM}\cup{\cal L}_{\rm SC}$ are lattice indices, $\lambda$ labels spin and orbital degrees of freedom, and $\tau$ corresponds to the particle-hole degree of freedom. The equilibrium low-energy physics of the SM wire can be extracted from the block labeled by indices associated with SM degrees of freedom. More specifically, we define the {\em effective} Green function of the SM wire as $[G_{\rm eff}(\omega)]_{mn} = [G(\omega)]_{mn}$
for all $m=\{i,\alpha,\tau\}$, $n=\{j,\beta,\tau^\prime\}$ with $i,j\in{\cal L}_{\rm SM}$. The effective Green function can be expressed in terms of $H_{\rm SM}$, $H_{\rm ext}$, the superconducting Green function $G_{\rm SC}(\omega) = (\omega - H_{\rm SC})^{-1}$, and the coupling matrix $\widetilde{T}$. Explicitly, we have \cite{Sau2010,Potter2011,Stanescu2013}
\begin{equation}
G_{\rm eff}(\omega) = \left[\omega - H_{\rm SM}-H_{\rm ext} -\Sigma(\omega)\right]^{-1},            \label{Geff}
\end{equation}
where $\Sigma(\omega) = (T\otimes \tau_z)~\! G_{\rm SC}(\omega)~\! (T^\dagger\otimes \tau_z)$ is an interface self-energy that incorporates the effect of the coupling to the superconductor. Here, $\tau_z$ is a Pauli matrix associated with the particle-hole degree of freedom and all quantities in Eq. (\ref{Geff}) are matrices indexed by the set of parameters $\{i, \alpha, \tau\}$.
Working with $G_{\rm eff}$ corresponds to `integrating out' the SC degrees of freedom. However, we emphasize that deriving Eq. (\ref{Geff}) involves {\em no approximation}, hence the low-energy physics described by $G_{\rm eff}$ is exactly the {\em same} as that described by the Hamiltonian in Eq. (\ref{Htot}).  

For this study we use the generic Green function framework defined by Eq. (\ref{Geff}).
Since our focus is the basic understanding of the proximity-induced renormalization of the low-energy physics, we model the system using simple tight-binding Hamiltonians. This involves a relatively small number of independent parameters, thus simplifying the analysis while completely capturing the basic physics.  
Specifically, we model the semiconductor wire (including the effects of external fields) as a set of $N_y$ coupled parallel chains described by 
\begin{eqnarray}
H_{\rm SM} + H_{\rm ext}  &=&  -\sum_{{\bm i},{\bm \delta}} t_{sm}^\delta \left(c^{\dagger}_{\bm i}c_{{\bm i}+{\bm \delta}} +c^{\dagger}_{{\bm i}+{\bm \delta}}c_{\bm i}\right) \nonumber \\
&+& \frac{i}{2}\sum_{{\bm i},{\bm \delta}}\alpha_R^\delta \left(c^{\dagger}_{{\bm i}+\delta_x}\hat{\sigma}_y c_{{\bm i}}-c^{\dagger}_{{\bm i}+{\delta_y}}\hat{\sigma}_x c_{\bm i} - {\rm h}.{\rm c}.\right)  \nonumber\\
&+& \sum_{i}(V_{\bm i}-\mu)c^{\dagger}_{\bm i}c_{\bm i} + \Gamma\sum_{\bm i}c^{\dagger}_{\bm i} \hat{\sigma}_x c_{\bm i}, \label{Hsm}
\end{eqnarray}
where $\hat{\sigma}_\mu$, with $\mu=x, y, z$, are Pauli matrices. The position is labeled by ${\bm i}=(i_x, i_y)$, with $1\leq i_y\leq N_y$ designating the chain and $1\leq i_x\leq N_x$ representing the position along the chain. The nearest neighbors corresponding to the longitudinal and transverse directions are given by $\delta_x=\pm 1$ and $\delta_y=\pm 1$, respectively. The matrix elements for hopping along and across the chains are 
$t_{sm}^{\delta_x}=t_0$ and $t_{sm}^{\delta_y}=t_0^\prime$, respectively, and the chemical potential of the wire is $\mu$. We note that the purely one-dimensional case, $N_y=1$, is a situation extensively considered in the earlier literature. Our use of a multi-chain ($N_y>1$) model allows multiband and interband-coupling effects to be included in the theory in a straightforward manner. In fact, the corresponding bands represent the confinement-induced 1D subbands, which are determined by the geometry of the wire cross section and the effective confinement potential in the nanowire.
The second term in Eq. (\ref{Hsm}) describes the Rashba spin-orbit coupling and  the strengths of the longitudinal and transverse components of this coupling are characterized by the coefficients $\alpha_R^{\delta_x}=\alpha_{R}$ and $\alpha_R^{\delta_y}=\alpha_R^\prime$, respectively. 
The position-dependent local term $V_{\bm i}$ describes the effective electrostatic potential generated by external gates and by the charge inside the wire.  Finally, the Zeeman splitting corresponding to a magnetic field applied along the wire is characterized  by the parameter $\Gamma$.

The superconductor is modeled at the mean-field level using the Bogoliubov - de Gennes formalism (BdG) and is assumed to be a bulk system, while the SM-SC interface is assumed to be planar. Within these approximations, the effect of the superconductor is captured by a local self-energy proportional to the surface Green function of the superconductor \cite{Stanescu2010,Stanescu2013}, 
\begin{equation}
\Sigma_{{\bm i}{\bm i}^\prime}(\omega) = -\delta_{{\bm i}{\bm i}^\prime}\widetilde{\gamma}_{\bm i} \left[\frac{\omega+\Delta_0\hat{\sigma}_y\hat{\tau}_y}{\sqrt{\Delta_0^2-\omega^2}}+\zeta\hat{\tau}_z\right], \label{Sigma}
\end{equation}
where $\widetilde{\gamma}_{\bm i} = \tilde{t}_{\bm i}^2\nu_F$ is the effective SM-SC coupling, with $\tilde{t}_{\bm i}$ being the nearest-neighbor hopping across the SM-SC interface at position ${\bm i}$ and $\nu_F$ the  surface density of states of the SC metal at the Fermi energy  \cite{Stanescu2010}. The Pauli matrices $\hat{\sigma}_\mu$ and $\hat{\tau}_\mu$ are associated  with  the spin and Nambu spaces,  respectively,  and $\zeta$ represents  a proximity-induced  shift  of  the  chemical  potential.  Below, we take $\zeta = 0$, for simplicity (as well as to minimize the number of unknown parameters).

We emphasize that, while the framework defined by Eq. (\ref{Geff}) is quite general, the self-energy (\ref{Sigma}) and the the Hamiltonian (\ref{Hsm}) contain  significant simplifying assumptions. In particular, the self-energy does not include non-local contributions $\Sigma_{{\bm i}{\bm i}^\prime}$ with ${\bm i}\neq{\bm i}^\prime$ and, more importantly, does not incorporate Coulomb and finite size effects (e.g., due to finite thickness in thin superconducting islands), as well as effects due to the presence of disorder and finite magnetic fields. These effects are expected to have important consequences in real devices and should be taken into account for a quantitative comparison with experiment. However, since these are secondary effects in the context of the proximity-induced renormalization, we do not discuss them explicitly. In addition, including some these non-essential complications in the theory has, at this stage, some degree of arbitrariness, since the experimentally-relevant parameters describing them are simply not known.

\section{Infinite wires: induced gap and phase diagram} \label{IW}

Consider now an infinitely-long, uniform system described by Eqs. (\ref{Geff}-\ref{Sigma}). After Fourier transforming with respect to $i_x$, the effective Green function takes the form
\begin{eqnarray}
&~&G_{\rm eff}^{-1}(\omega, k) = \omega\left(1+\frac{\hat{\tilde{\gamma}}}{\sqrt{\Delta_0^2-\omega^2}}\right) + \frac{\Delta_0\hat{\tilde{\gamma}}\hat{\sigma}_y\hat{\tau}_y}{\sqrt{\Delta_0^2-\omega^2}}~~~~~~ \label{Ginv} \\
&~&-\left(\xi(k) -t_0^\prime\hat{\lambda}_x + \Gamma\hat{\sigma}_x +\alpha_R \sin k~\! \hat{\sigma}_y\right)\hat{\tau}_z +\frac{1}{2}\alpha_R^\prime\hat{\lambda}_y\sigma_x, \nonumber
\end{eqnarray} 
where $\xi(k) = -2 t_0 \cos k -\mu$ is the (single band) energy dispersion in the absence of spin-orbit coupling, the coupling matrix $\hat{\tilde{\gamma}}$ has matrix elements $\delta_{i j}\tilde{\gamma}_j$, and the matrices $\hat{\lambda}_x$ and $\hat{\lambda}_y$ of dimension $N_y\times N_y$ have the form
\begin{equation}
\hat{\lambda}_x = \left(
\begin{array}{cccc}
0 & 1 & 0 & \dots \\
1 & 0 & 1 & \dots \\
0 & 1 & 0 & \dots \\
\dots \!\! & \!\dots\!\! & \!\dots\!\! & \!\dots 
\end{array}\right),~~~~~~~
\hat{\lambda}_y = \left(
\begin{array}{cccc}
0 & \!-i\! & 0 & \dots \\
i & 0 & \!-i\! & \dots \\
0 & i & 0 & \dots \\
\dots \!\! & \!\dots\!\! & \!\dots\!\! & \!\dots 
\end{array}\right),
\end{equation}
if $N_y\geq 2$ and $\hat{\lambda}_x = \hat{\lambda}_y =0$ for $N_y=1$.
To simplify the notation, we omit all unit matrices in Eq. (\ref{Ginv}). The superconducting proximity effect is accounted for through the frequency-dependent terms proportional to the coupling matrix  $\hat{\tilde{\gamma}}$. Note that within the local self-energy approximation this matrix is diagonal, but, in general, off-diagonal terms may be present. The dependence of  $\tilde{\gamma}_j$ on the chain index models the realistic situation corresponding to a semiconductor wire partially covered by the s-wave superconductor. The low-energy states of the hybrid system are given by the poles of the Green function and can be obtained by solving the equation
\begin{equation}
\det[G_{\rm eff}(\omega, k)] = 0.  \label{poles}
\end{equation}

\subsection{Single band approximation}

The simplest case corresponds to a system characterized by confinement-induced nanowire bands that are well-separated, i.e., the inter-band gaps are large compared to the SM-SC effective coupling, $t_0^\prime \gg \tilde{\gamma}_j$. In this case, the bands can be treated as independent and we can model a generic band using a single-chain model, i.e., Eq. (\ref{Ginv}) with $N_y=1$. Without the renormalization corrections, this is the nanowire model that was extensively used so far in the literature for studying the Majorana modes in hybrid SC-SM structures. With the self-energy contribution due to the parent superconductor,  the energies of the low-lying states with $k=0$ can be obtained from Eq. (\ref{poles}), which reduces to 
\begin{equation}
\omega\left(1+\frac{\tilde{\gamma}}{\sqrt{\Delta_0^2-\omega^2}}\right) = (\pm) \sqrt{\bar{\mu}^2 + \frac{\tilde{\gamma}^2\Delta_0^2}{\Delta_0^2-\omega^2}}~\pm~\Gamma,   \label{poles1}
\end{equation}
where the signs of the two right hand side terms are independent and $\bar{\mu}=\xi(0)$, i.e., the chemical potential measured from the bottom of the band (in the absence of spin-orbit coupling).

%%%%%%%%%%%%%%%%%%%%%%%%%%%%
\begin{figure}[t]
\begin{center}
\includegraphics[width=0.48\textwidth]{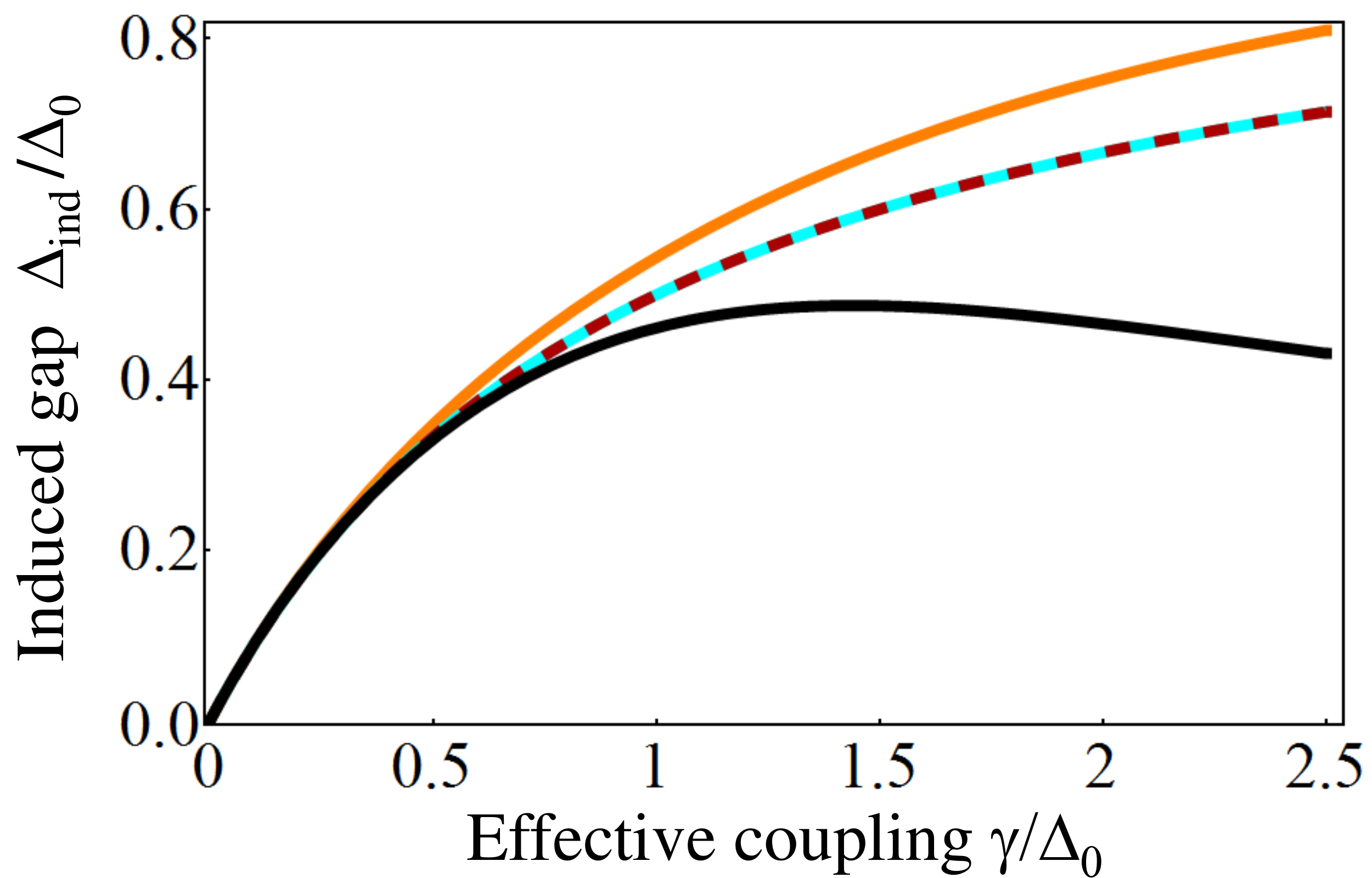}
\vspace{-3mm}
\end{center}
\caption{(Color online) Dependence of the induced gap on the effective SM-SC coupling. The dashed line corresponds to the weak-coupling expression $\Delta_{\rm ind}=\tilde{\gamma}\Delta_0/(\tilde{\gamma}+\Delta_0)$, while the orange (gray) line is obtained by solving Eq. (\ref{Dind}). The black line is the induced gap for a two-band system with the SM-SC coupling matrix given by Eq. (\ref{gmb}), inter-band spacing $\Delta\epsilon=2$ meV, and no transverse spin-orbit coupling.}
\vspace{-2mm}
\label{Fig01}
\end{figure}
%%%%%%%%%%%%%%%%%%%%%%	

The topological quantum phase transition (TQPT) between the trivial and topological superconducting phases is characterized by the vanishing of the quasiparticle gap at $k=0$. Consequently, we can determine the critical Zeeman field $\Gamma_c$ at the transition by looking for the solution of Eq. (\ref{poles1}) with $\omega=0$. We have
\begin{equation}
\Gamma_c = \sqrt{\bar{\mu}^2+\tilde{\gamma}^2}.  \label{Gmmc}
\end{equation}
Note that Eq. (\ref{Gmmc}) is similar to the ``standard'' expression of the critical field, $\Gamma_c = \sqrt{\bar{\mu}^2+\Delta_{\rm ind}^2}$, where $\Delta_{\rm ind}$ is the so-called {\em induced gap}. However, the minimum critical field, which obtains at $\bar{\mu}=0$, is determined by the effective semiconductor-superconductor coupling $\tilde{\gamma}$, rather than the induced gap, $\Delta_{\rm ind}$.  We mention that in general these two quantities are different, except in the very weak coupling limit (see below).

Formally, we define the {\em induced gap} $\Delta_{\rm ind}(\mu)$ as the minimum with respect to $k$ of the lowest energy quasiparticle state given by Eq. (\ref{poles}) at $\Gamma = 0$. For a single band, this quantity is practically  independent of the chemical potential for values of $\mu$ above the bottom of the band. We denote this constant as $\Delta_{\rm ind}$.  
One can show that $\Delta_{\rm ind}$ can be obtained by setting $\bar{\mu}=0$ in Eq. (\ref{poles1}). Using this observation,  we determine the induced gap by solving the equation
\begin{equation}
\Delta_{\rm ind}\sqrt{\Delta_0+\Delta_{\rm ind}} = \tilde{\gamma}\sqrt{\Delta_0-\Delta_{\rm ind}}.  \label{Dind}
\end{equation} 
In the weak coupling limit, $\Delta_{\rm ind}\ll \Delta_0$, we obtain $\Delta_{\rm ind} = \tilde{\gamma}\Delta_0/(\tilde{\gamma}+\Delta_0)\approx \tilde{\gamma}$ and the equation for the critical field reduces to the ``standard'' form. In general, however, $\Delta_{\rm ind} <\tilde{\gamma}$ and the critical Zeeman field is generically {\em larger} than the value predicted based on the induced gap. The numerical solution of Eq. (\ref{Dind}) for arbitrary coupling strength is shown in Fig. \ref{Fig01}.

From Eq. (\ref{poles1}) we notice that the quasiparticle gap goes linearly to zero with the applied magnetic field as one approaches the TQPT. The effective g-factor can be determined from the corresponding slope. However, we emphasize that the slope $g_{\rm eff}=\frac{1}{\mu_B}|d\omega/dB|$, where $B$ is the magnetic field and $\mu_B$ the Bohr magneton, represents the proximity-renormalized g-factor, rather than the ``bare'' g-factor, $g_0$. Explicitly, we have
\begin{equation}
g_{\rm eff} = \frac{g_0}{1+\tilde{\gamma}/\Delta_0}.
\end{equation}  
In the intermediate ($\tilde{\gamma}\sim \Delta_0$) and strong ($\tilde{\gamma} \gg \Delta_0$) coupling regimes the effective g-factor has values that are significantly smaller than the bare semiconductor value $g_0$. This has obvious experimental implications since the TQPT is defined by the spin splitting [see Eq. (\ref{Gmmc})], which connects with the experimental tuning parameter -- the magnetic field -- through the renormalized g-factor.

\subsection{Multi-band systems}

In the previous section we have shown that for a one-dimensional system the phase boundary is determined by the chemical potential and the effective  SM-SC coupling, being independent of the bulk gap and the spin-orbit coupling, while the induced gap is determined by $\Delta_0$ and $\tilde{\gamma}$, again, being $\alpha_R$-independent. In the multi-band case the situation is, in general different and more complex. The difference arises from the presence of an additional energy scale: the inter-band spacing, which is represented in our model by the inter-chain hopping $t_0^\prime$. In addition, the spin-orbit coupling acquires a transverse component represented in Eq. (\ref{Ginv}) by the last term (i.e., the term proportional to $\alpha_R^\prime$). However, we emphasize that this Rashba-type spin-orbit coupling is a simplified effective model that may not capture all aspects of this phenomenon and a more detailed modeling may be necessary for quantitative predictions. Such a detailed modeling, which is  non-generic and depends crucially on all the specific features characterizing the nanowire and its environment, is beyond the scope of the current study.

We distinguish two basic limits: i) The independent band regime, which is realized in systems with large inter-band spacing, $t_0^\prime \gg \tilde{\gamma}$, or systems characterized by a SM-SC effective coupling that does not mix different bands, e.g., a matrix $\hat{\tilde{\gamma}}$ in Eq. (\ref{Ginv}) that is diagonal. Note that the symmetry of the SM-SC coupling is not generic, hence in practice the independent band regime is expected to be relevant only in systems with well-separated bands, typically at very low occupancy. ii) The coupled band regime, which corresponds to  $t_0^\prime \sim \tilde{\gamma}$ and an arbitrary spatial profile of the SM-SC coupling. Note that the spectrum of a semiconductor wire is characterized by sets of nearly-degenerate higher-energy  bands. The coupled band regime is expected to become relevant when the chemical potential is in the vicinity of such nearly-degenerate bands.

To illustrate the main features associated with the two regimes described above, we consider a two-chain model, $N_y=2$, which corresponds to a system that has the chemical potential in the vicinity of two orbital bands that are well separated from the rest of the spectrum. Of course, a quantitative analysis requires the detailed modeling of the transverse degrees of freedom of the SM wire, but here we focus on the key aspects of the proximity-induced low-energy renormalization in the presence of generic inter-band coupling, which are clearly captured by the simplified model.   

The independent band regime is fully characterized by the results obtained for a single band. Indeed, let us consider Eq. (\ref{Ginv}) for $N_y=2$. For simplicity we neglect the transverse spin-orbit coupling and check numerically the accuracy of this approximation. It turns out that this is an excellent approximation for all reasonable values of  $\alpha_R^\prime$. The critical fields $\Gamma_{c1}$ and $\Gamma_{c2}$ corresponding to the low- and high-field boundaries of the topological superconducting phase are given by the following generalization of Eq. (\ref{Gmmc})
\begin{eqnarray}
\Gamma_{c1} &=& {\rm min}\left[\sqrt{(\mu - \epsilon_1)^2+\tilde{\gamma}^2}, ~\sqrt{(\mu - \epsilon_2)^2+\tilde{\gamma}^2}\right], \nonumber
\end{eqnarray}
\begin{eqnarray}
\Gamma_{c2} &=& {\rm max}\left[\sqrt{(\mu - \epsilon_1)^2+\tilde{\gamma}^2}, ~\sqrt{(\mu - \epsilon_2)^2+\tilde{\gamma}^2}\right],   \label{Gmmc12}
\end{eqnarray}
where $\epsilon_{1,2} = -2t_0 \mp t_0^\prime$ designate the bottoms of the two bands. We note that Eq. (\ref{Gmmc12}) can be conveniently expressed using the {\em band}, rather than the {\em chain} (i.e., real space) representation in terms of the effective couplings $\gamma_n$ associated with each band. In general, if $\tilde{\gamma}_{ij}$ is the position-dependent SM-SC effective coupling and $\phi_n(i)$ is the transverse component of a wave function associated with band $n$, we have
\begin{equation}
\gamma_{nm} = \langle \phi_n|\tilde{\gamma}|\phi_m\rangle.  \label{gnm}
\end{equation}
Generally, the matrix $\gamma_{nm}$ contains off-diagonal terms that couple different bands. However, for our  simple model this matrix is diagonal and we have $\gamma_1=\gamma_2=\tilde{\gamma}$, where $\gamma_n\equiv \gamma_{nn}$ designates a diagonal element. The generalization of Eq. (\ref{Gmmc12}) for a system with an arbitrary number of independent bands corresponds to a low-field phase boundary of the form
$\Gamma_{c1} = {\rm min}[\sqrt{(\mu-\epsilon_1)^2+\gamma_1^2}, ~\sqrt{(\mu-\epsilon_2)^2+\gamma_2^2}, ~\dots]$. 

In the independent band regime, the induced gap associated with band $n$ is given by Eq. (\ref{Dind}) with $\tilde{\gamma}\rightarrow \gamma_n$. In general, different bands are characterized by different values of the SM-SC coupling, hence by different values of the induced gap. The minimum gap is determined by the occupied band with the lowest value of $\gamma_n$, i.e., the weakest coupled band. 

Next, we consider the coupled band regime, which is characterized by nonzero values of the off-diagonal effective coupling $\gamma_{nm}$ and small inter-band spacings. We focus on two specific forms of the SM-SC coupling that  correspond to (a) the SC being coupled to only one SM chain and (b) the SM-SC coupling having non-local components. The corresponding coupling matrices in the chain ($\hat{\tilde{\gamma}}$) and band ($\hat{\gamma}$) representations are:
\begin{eqnarray}
(a)~~~~\hat{\tilde{\gamma}}&=&\left(\begin{array}{cc} 2 & 0 \\ 0 & 0\end{array}\right)\gamma, 
~~~~~~~~~~\hat{\gamma}=\left(\begin{array}{cc} 1 & 1 \\ 1 & 1\end{array}\right)\gamma,   \label{gma} \\
(b)~~~~\hat{\tilde{\gamma}}&=&\left(\begin{array}{rr} 9 & \!\!-3 \\ \!\!-3 & 1\end{array}\right)\frac{\gamma}{2}, 
~~~~~~\hat{\gamma}=\left(\begin{array}{cc} 1 & 2 \\ 2 & 4\end{array}\right)\gamma.   \label{gmb}
\end{eqnarray}
Note that all these matrices depend on a single parameter, $\gamma$, which characterizes the strength of the SM-SC coupling, and that both scenarios involve a proximity-induced coupling between the two bands, i.e., $\gamma_{12}\neq 0$. By comparison, the independent band regime discussed above corresponds to $\hat{\tilde{\gamma}} = \hat{{\gamma}} = \gamma I_2$, where $I_2$ is the $2\times 2$ identity matrix. 

%%%%%%%%%%%%%%%%%%%%%%%%%%%%
\begin{figure}[t]
\begin{center}
\includegraphics[width=0.49\textwidth]{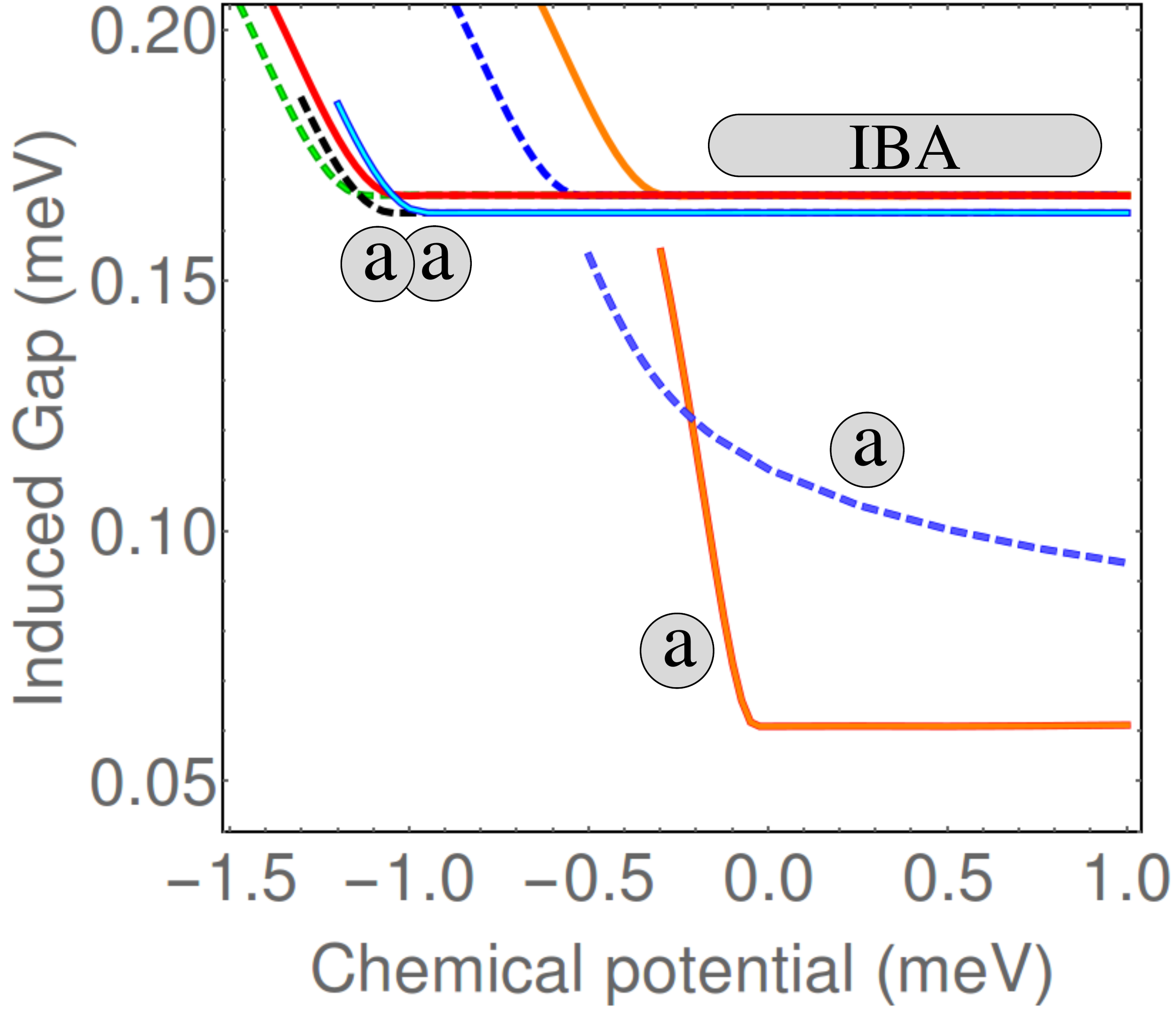}
\vspace{-3mm}
\end{center}
\caption{(Color online) Induced gap as function of the chemical potential for a system with $\Delta_0=0.25$ meV and $\gamma = 0.375$ meV. The full lines are calculated in the absence of transverse spin-orbit coupling, while the dashed lines correspond to $\alpha_R^\prime=0.9$ meV. The curves with an upturn below $\mu=-1$ meV correspond to an inter-band spacing $\Delta\epsilon = 2$ meV (with the bottom of the lowest band at $\epsilon_1=-1$ meV), while the curves with an upturn at higher values of $\mu$ are for  $\Delta\epsilon = 0.5$ meV ($\epsilon_1=-0.25$ meV). The top four curves are generated within the independent band approximation (IBA) with a coupling matrix $\hat{\gamma}=\gamma I_2$, while the other four lines correspond to scenario (a) with the coupling matrix given by Eq. (\ref{gma}). The $\mu$-independent value of $\Delta_{\rm ind}$ for the top four curves is given by Eq. (\ref{Dind}).}
\vspace{-2mm}
\label{Fig02}
\end{figure}
%%%%%%%%%%%%%%%%%%%%%%

The induced gap $\Delta_{\rm ind}(\mu)$ corresponding to scenario (a) is shown in Fig. \ref{Fig02} for different values of the inter-band spacing $\Delta\epsilon = \epsilon_2-\epsilon_1 = 2t_0^\prime$ and different transverse spin-orbit couplings. For comparison, we also show the corresponding gap in the independent band regime. Focusing first on the independent band approximation (IBA, top four curves), we notice that the constant value of the induced gap at large $\mu$ is given by Eq. (\ref{Dind}), i.e., $\Delta_{\rm ind}=0.167$ meV, independent of the inter-band splitting $\Delta\epsilon$ and the transverse Rashba coefficient $\alpha_R^\prime$. The only feature that depends on these parameters is the upward turn in $\Delta_{\rm ind}(\mu)$ with decreasing chemical potential. This upturn corresponds to the chemical potential getting below the bottom of the lowest-energy band, i.e., the system becoming completely depleted. The change in the location of this upturn merely reflects the dependence of the energy of the bottom of the lowest band on $\epsilon_1$ and  $\alpha_R^\prime$. Finally, we note that the slope of the upturn, which is (minus) one in the weak-coupling limit, $\Delta_{\rm ind}(\mu) \approx -\bar{\mu}$, becomes strongly renormalized in the intermediate-coupling regime ($\gamma = 1.5\Delta_0$) studied here.

Next, we discus scenario (a), which corresponds to one SM chain being coupled to the superconductor.  This is a crude model of a semiconductor wire half-covered by a superconductor, such as the systems recently studied experimentally in Copenhagen using InAs/Al core-shell-nanowire structures \cite{Chang2014,Krogstrup2015,Albrecht2016,Deng2016}.
  If the inter-band spacing is large, $\Delta\epsilon \gg \gamma$, the system behaves qualitatively the same as in the independent band approximation. This is illustrated by the top two curves marked by an (a) in Fig. \ref{Fig02}. The only difference is that the $\mu$-independent  $\Delta_{\rm ind}$ is slightly lower than the value given by Eq. (\ref{Dind}). The situation is completely different when $\Delta\epsilon$ and  $\gamma$ become comparable, as illustrated by the lower two curves in Fig. \ref{Fig02}. Most notable are the drastic drop of the induced gap and the strong dependence on the transverse Rashba coupling. Intuitively, we can understand these effects as a result of the proximity-induced coupling of two nearly-degenerate  bands. The ``bare'' transverse modes associated with the two bands, which have the same coupling to the SC, as expressed by Eq. (\ref{gma}), are not directly related to the eigenstates of the composite system, since $\hat{\gamma}$ has off-diagonal coupling elements. We can introduce  new ``perturbed  modes'' that are weakly coupled to each other. However, one of these modes is strongly coupled to the SC, while the other is characterized by a weaker effective coupling. In the degenerate band limit ($t_0^\prime \rightarrow 0$), for example, the ``perturbed modes'' are localized on chain `1' (with effective coupling $\gamma_1=2\gamma$ and chain `2' (with effective coupling $\gamma_2=0$), respectively.
The weakly coupled mode is responsible for the lower values of the induced gap. In addition, the ``perturbed modes'' depend on the transverse spin-orbit coupling, which explains the strong dependence on $\alpha_R^\prime$ in this regime. 

We have argued that in the coupled band regime, which is  characterized by the presence of proximity-induced inter-band coupling and inter-band spacings comparable to the effective SM-SC coupling, the low-energy physics of the hybrid system is controlled by a renormalized ``perturbed'' mode that has a weaker effective SM-SC coupling than the ``bare'' modes. To illustrate the fact that this mechanism is rather generic, we calculate the induced gap as function of $\gamma$ for scenario (b) given by Eq. (\ref{gmb}). The result is shown in Fig. \ref{Fig01} (black curve). Increasing the SM-SC coupling drives the system into the coupled band regime, which results in the emergence of a renormalized mode characterized by a weaker effective SM-SC coupling. As a result, the induced gap decreases with increasing $\gamma$ because of the stronger renormalization by the SC (in contrast to the bare theory, where the induced gap increases with increasing coupling), which appears puzzling. However, this feature can be naturally understood within the weakly-coupled renormalized mode framework described above. 

Let us now turn our attention to analyzing the phase diagram. As a general remark, we note that a phase boundary corresponds to poles of the effective Green function given by Eq. (\ref{Ginv}) at $k=0$ and $\omega=0$. Consequently, the phase boundaries are independent of the longitudinal Rashba coefficient $\alpha_R$ and the bulk SC gap $\Delta_0$. By contrast, the induced gap is manifestly dependent on $\Delta_0$, hence we do not expect any direct relation between $\Delta_{\rm ind}$ and the critical Zeeman field, except in the weak coupling limit where $\Delta_{\rm ind}\approx \gamma$.  
The independent band regime is described by Eq. (\ref{Gmmc12}) and its generalizations, as discussed above. In the coupled band regime, the low-field boundary of the topological phase  is given by
\begin{eqnarray}
\Gamma_{c1}(\mu) &=& {\rm min}\left[\sqrt{\beta(\mu)-\sqrt{\beta^2(\mu)-\alpha(\mu)}}, \right. \nonumber \\ 
&~&~~~~~~~~\left.\sqrt{\beta(\mu)+\sqrt{\beta^2(\mu)-\alpha(\mu)}}\right], \label{Gmmc1a}
\end{eqnarray}
with,
\begin{eqnarray}
\alpha(\mu)&=& \gamma_{12}^4+[\gamma_{11}^2+(\epsilon_1-\mu)^2][\gamma_{22}^2+(\epsilon_2-\mu)^2] \nonumber \\
&-&2\gamma_{12}\left[\gamma_{11}\gamma_{22}-(\epsilon_1-\mu)(\epsilon_2-\mu)\right],  \\
\beta(\mu)&=& \gamma_{11}^2+2\gamma_{12}^2+ \gamma_{22}^2+(\epsilon_1-\mu)^2+(\epsilon_2-\mu)^2, \nonumber
\end{eqnarray}
where $\gamma_{nm}$ are the elements of the coupling matrix $\hat{\gamma}$ and $\epsilon_n$ is the energy at the bottom of the $n$-th band. These equations were derived in the absence of transverse spin-orbit coupling. If $\alpha_R^\prime\neq 0$ the phase diagram can be obtained by solving Eq. (\ref{poles}) numerically and determining the Zeeman fields corresponding to zero-energy solutions. Note that the high-field boundary of the topological phase is given by an equation similar to Eq. (\ref{Gmmc1a}), but with ``min'' $\rightarrow$ ``max''.

%%%%%%%%%%%%%%%%%%%%%%%%%%%%
\begin{figure}[t]
\begin{center}
\includegraphics[width=0.48\textwidth]{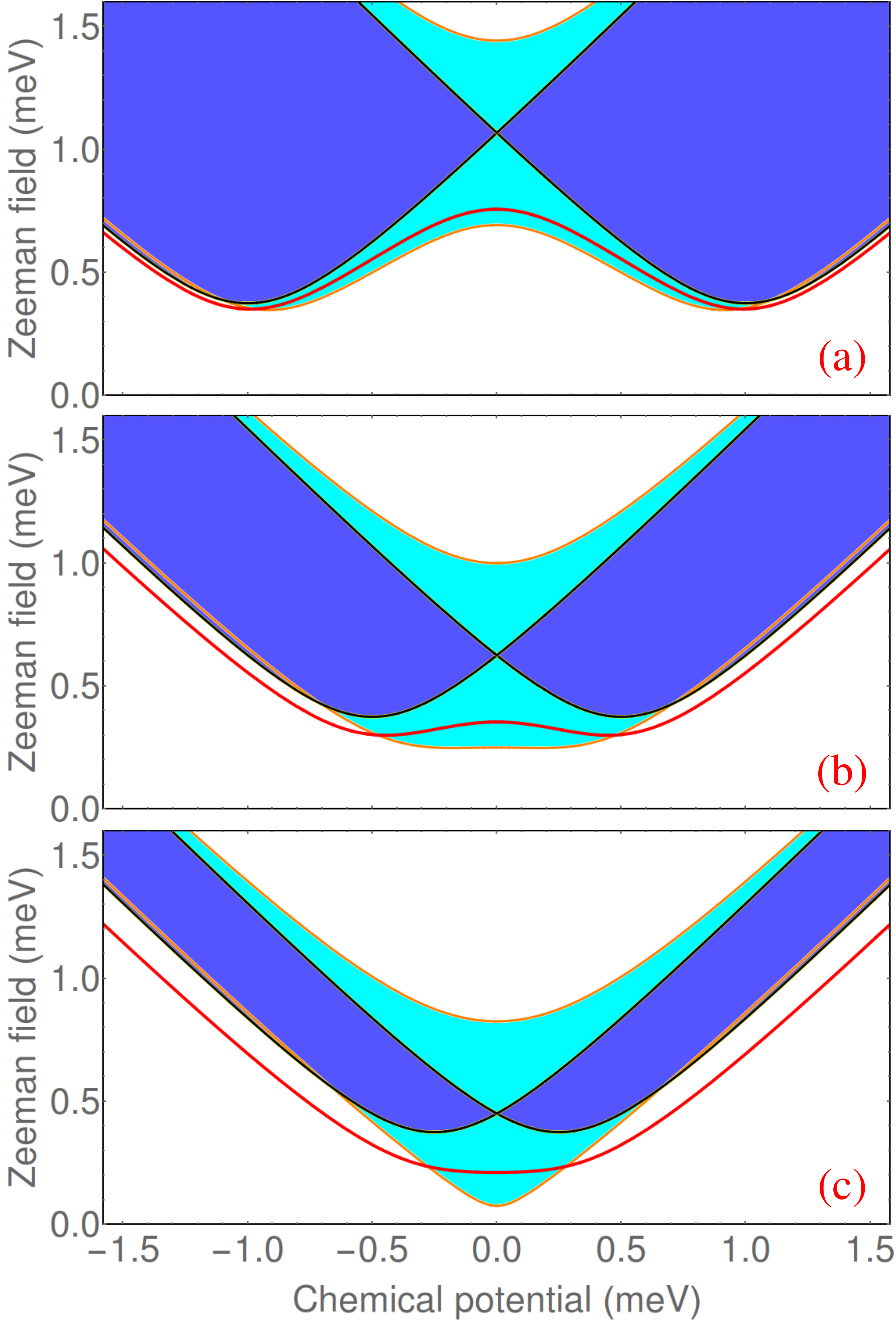}
\vspace{-2mm}
\end{center}
\caption{(Color online) Topological phase diagram of a two-band system with $\gamma=0.375$, $\alpha_R^\prime=0$, and three different values of the inter-band spacing: (a) $\Delta\epsilon=2$ meV, (b) $\Delta\epsilon=1$ meV, and (c) $\Delta\epsilon=0.5$ meV. The blue (dark gray) areas represent the topological superconducting phase in the independent band regime. In the coupled band regime corresponding to Eq. (\ref{gma}), i.e., scenario ``a'' , the topological phase extends into the cyan (light gray) areas. In the presence of transverse spin-orbit coupling with $\alpha_R^\prime=0.75$ meV the low-field phase boundary is given by the thick red line.}
\vspace{-2mm}
\label{Fig03}
\end{figure}
%%%%%%%%%%%%%%%%%%%%%%	

The dependence of the phase boundaries on the inter-band spacing for scenario (a) is shown in Fig. \ref{Fig03}.  The corresponding phase boundaries  in the independent phase approximation are also shown, for comparison. First, we note that the most significant difference between the two cases occurs when the chemical potential is above the bottom of the first band but does not touch the second band, i.e., $\epsilon_1 <\mu < \epsilon_2$. In this region, characterized in the independent band approximation by a point associated with a sub-band crossing where the width of the topological region vanishes, adding inter-band coupling results in a significant expansion of topological superconductivity (cyan/light gray areas in Fig. \ref{Fig03}). If the bands are well separated, $\Delta\epsilon \gg \gamma$, the additional topological SC region occurs at large values of the Zeeman field and might not be experimentally observable. By contrast, for small separations [see panels (b) and (c)] the phase boundary extends toward $\Gamma=0$ and proximity-induced band coupling becomes highly relevant. As discussed above, we can intuitively understand this behavior in terms of an emerging ``renormalized mode'' that is weakly coupled to the bulk superconductor and, consequently, determines a topological phase transition  at low values of the Zeeman field. 

%%%%%%%%%%%%%%%%%%%%%%%%%%%%
\begin{figure}[t]
\begin{center}
\includegraphics[width=0.48\textwidth]{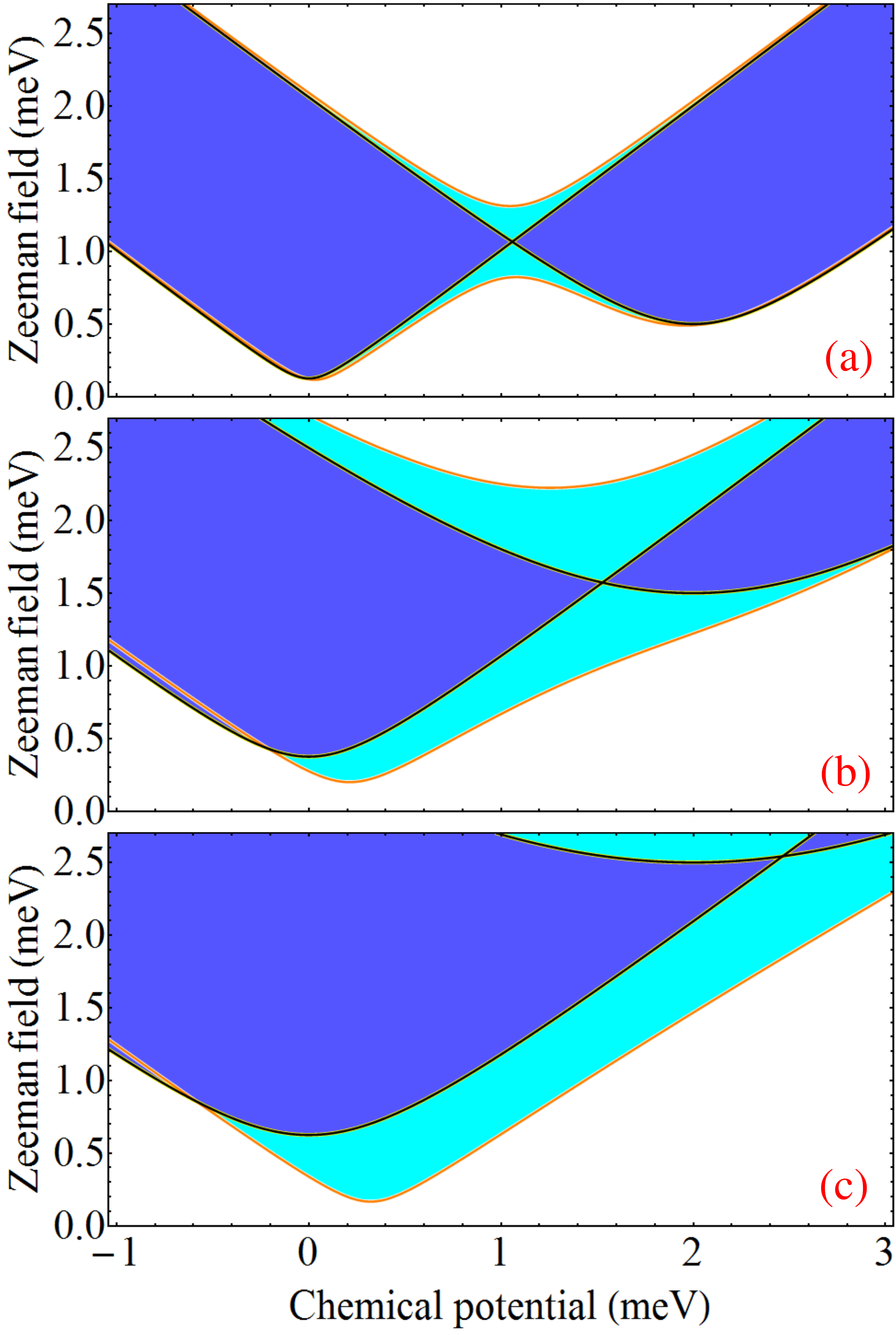}
\vspace{-2mm}
\end{center}
\caption{(Color online) Topological phase diagram of a two-band system with $\Delta\epsilon=2$ meV, $\alpha_R^\prime=0$, and three different values of the SM-SC coupling strength: (a) $\gamma=0.5$ meV, (b) $\gamma=1.5$ meV, and (c) $\gamma=2.5$ meV.  The blue (dark gray) regions represent the topological superconducting phase in the independent band approximation with $\gamma_1=\gamma_2/4=\gamma$.  In the coupled band regime corresponding to Eq. (\ref{gmb}), i.e., scenario ``b'' , the topological phase extends into the cyan (light gray) areas.}
\vspace{-2mm}
\label{Fig04}
\end{figure}
%%%%%%%%%%%%%%%%%%%%%%	

Typically in the literature it is argued that the phase boundaries do not depend on the strength of the spin-orbit coupling. Strictly speaking, this is only true for the longitudinal component of the spin-orbit coupling. The transverse component, on the other hand, impacts the location of the phase boundary, as shown in Fig. \ref{Fig03}. The red line corresponds to the low-field phase boundary of a system with coupling matrix given by Eq. (\ref{gma}) (i.e., scenario ``a'') and $\alpha_R^\prime=0.75$ meV. Note that for large values of the inter-band gap the change is small, but it becomes significant with reducing the band separation $\Delta\epsilon$.   Also note that the phase boundary is modified at all values of the chemical potential. The effect is equivalent to shifting the two bands so that $\epsilon_1\rightarrow \epsilon_1 - \delta$ and $\epsilon_2\rightarrow \epsilon_2 + \delta$, where $\delta$ is a positive quantity. This effect was already discussed in the context of Fig. \ref{Fig02}, where the shift of $\epsilon_1$ becomes manifest as a change in the location of the upturn of $\Delta_{\rm ind}(\mu)$. 

We have shown that the topological phase  boundaries of a system with proximity-induced coupled bands is significantly different from the phase  boundaries in the absence of this coupling, particularly at low values of the Zeeman field. Our analysis suggests that this effect becomes significant when the inter-band separation is comparable with the SM-SC coupling strength $\gamma$. To confirm this conclusion, we consider a system with fixed band separation and different values of the SM-SC coupling strength. Also, to show that the effects discussed above are not dependent on the bands being equally coupled to the SC, we focus on scenario ``b'' from Eq. (\ref{gmb}). Note that in the corresponding independent band approximation the upper band has an effective SM-SC coupling four times larger than the lower band, $\gamma_2=4\gamma_1=4\gamma$. The phase diagrams are shown in Fig. \ref{Fig04}. One can easily observe all the qualitative features discussed above. In particular, when the effective coupling strength becomes comparable with the inter-band separation the topological phase boundary extends into the low Zeeman field region, a phenomenon that can be naturally interpreted in terms of an emerging weakly-coupled ``renormalized band''. 

We conclude this section with a brief summary of the main results. For the induced gap, we distinguish a low-coupling regime and an intermediate/strong coupling regime based on the relative strengths of the SM-SC coupling and bulk SC gap. For $\gamma<0.5 \Delta_0$ the system is in the weak coupling regime and the induced gap is well approximated by the standard expression $\Delta_{\rm ind} = \gamma\Delta_0/(\gamma+\Delta_0)$. At larger couplings we need to also consider the effects of proximity-induced inter-band coupling. In the independent band regime $\Delta_{\rm ind}$ is well described by the single-band approximation given by Eq. (\ref{Dind}), while in the coupled band regime the effective coupling is strongly renormalized and $\Delta_{\rm ind}$ has a non-monotonic dependence on $\gamma$.  The phase diagram is independent of $\Delta_0$,  but depends strongly on the ratio between the effective coupling and the inter-band spacing. Again, the independent band regime is well described by the single band approximation [see Eq. (\ref{Gmmc12})], while in the coupled band regime the topological phase can extend significantly, most importantly at low values of the Zeeman field [see Eq. (\ref{Gmmc1a})]. Note that, in principle, the proximity-induced inter-band coupling may be important even in the weak-coupling regime, as long as as the effective SM-SC coupling is comparable with the inter-band spacing, $\gamma \approx \Delta\epsilon \ll \Delta_0$. 
The existence of nearly degenerate semiconductor bands makes the coupled band regime relevant in systems with high occupancy. 

\section{Finite wires: quasiparticle gap and Majorana splitting}

The analysis of the infinite wire has shown that the strength of the coupling between the semiconductor and the superconductor has a profound impact on both the induced gap and the phase diagram. However, our main focus is to understand the effect of this coupling on Majorana bound states, particularly on the energy splitting associated with overlapping Majoranas and the quasiparticle gap that protects these modes. To this end, we turn our attention to finite systems, which can host Majorana modes localized near the ends of the wire. 

For concreteness, we consider a relatively short wire of length $L_x=0.8$ $\mu$m, which in the weak coupling approximation corresponds to a system that hosts strongly overlapping Majorana modes at the two ends of the wire. The hybrid system is modeled as a set of parallel chains coupled to a superconductor (SC) using Eqns. (\ref{Geff}), (\ref{Hsm}), and (\ref{Sigma}). The values of model parameters used in the numerical calculations are: $t_0=1656.5/a_x^2$ meV, where $a_x$ is the lattice constant in nanometers (typically $a=10$ nm), $t_0^\prime=4.0$ meV (unless specified otherwise), $\alpha_R = 15.0/a_x$ meV, $\alpha_R^\prime= 0.9$ meV (unless specified otherwise), and $V_{\bm i}=0$. The value of $t_0$ corresponds to an effective mass $m^*=0.023 m_0$. 
We note that in the weak coupling limit $\bar{\gamma}_{\bm i}=\gamma \ll \Delta_0$ the induced gap is given by the effective coupling,  $\Delta_{\rm ind}\approx\gamma$, and the dynamical corrections are negligible,
$\gamma/\sqrt{\Delta_0^2-\omega^2} \ll 1$,  for all relevant energy values, $\omega <\Delta_{\rm ind}$.   In this limit, the system can be described in terms of an effective Bogoliubov -- de Gennes (BdG) Hamiltonian $H_{\rm eff} = H_{\rm SM}+H_{\rm ext} + \Sigma$, where the energy-independent interface self-energy is  
\begin{equation}
\Sigma_{{\bm i},{\bm i}^\prime} = - \Delta_{\rm ind}~\!\hat{\sigma}_y\hat{\tau}_y ~\!\delta_{{\bm i},{\bm i}^\prime}.
\end{equation}
We emphasize that most of the theoretical results discussed in the Majorana nanowire literature are based on this effective Hamiltonian approximation. The key question that we want to address is how relevant are these results for systems that do not satisfy the condition $\gamma \ll \Delta_0$. We note that in the experimentally-available SC-SM hybrid structures there is little control over the interface coupling and, therefore, the weak-coupling approximation is generically inapplicable. For example, the intensely studied epitaxial structures \cite{Krogstrup2015} are expected to be in the strong-coupling regime.

\subsection{Zeeman field dependence of the low-energy spectrum}

The low-energy spectrum of the SM-SC structure is determined numerically by solving the equation ${\rm det}[G_{\rm eff}(\omega)]=0$, where the effective Green function is a matrix given by Eq. (\ref{Geff}). First, we focus on the dependence of this spectrum on the applied Zeeman field  $\Gamma$ [see Eq. (\ref{Hsm})]. We note that, in general, the bulk superconductor is also affected by the presence of the external magnetic field. The effect can be naturally incorporated into the theory via the Green function of the SC, i.e., via $\Sigma(\omega)$. Here, we consider a very simplified model consisting of a self-energy given by Eq. (\ref{Sigma}) with a field-dependent bulk gap $\Delta_0(\Gamma)$. However, we emphasize that a better understanding  of the parent superconductor at finite magnetic fields in the presence of disorder remains  a major outstanding problem in this field. Addressing this problem is beyond the scope of the current work.

%%%%%%%%%%%%%%%%%%%%%%%%%%%%
\begin{figure}[t]
\begin{center}
\includegraphics[width=0.48\textwidth]{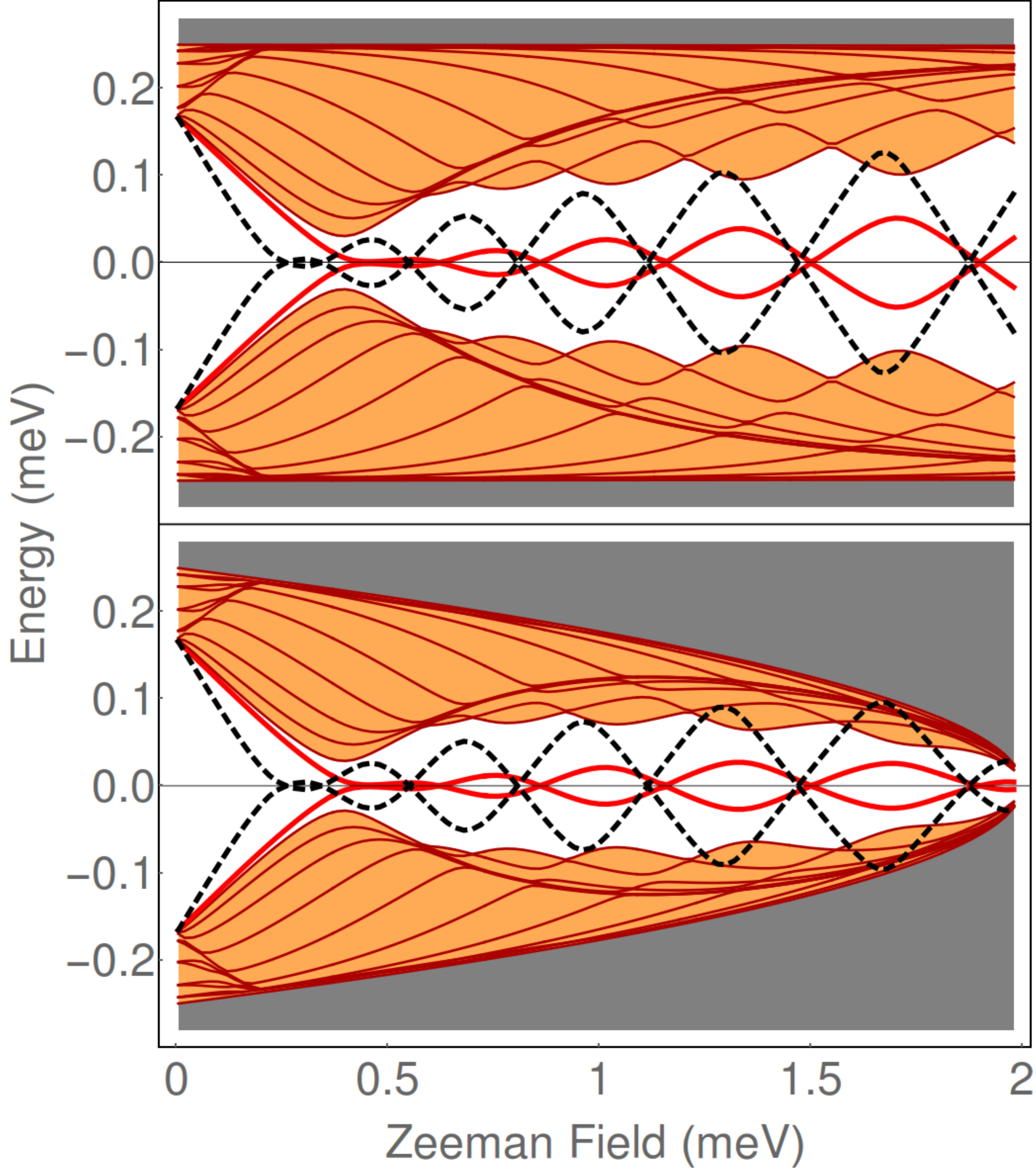}
\vspace{-3mm}
\end{center}
\caption{(Color online) Majorana splitting oscillations at weak coupling ($\Delta_0=0.82$ meV, $\gamma=0.25\Delta_0$ -- dashed black lines) and intermediate coupling ($\Delta_0=0.25$ meV, $\gamma=1.5\Delta_0$ -- red lines).  Both systems are characterized by an induced gap $\Delta\approx 0.167$ meV. The upper panel corresponds to a constant bulk SC gap, while the lower panel is for a field-dependent gap that vanishes at $\Gamma=2$ meV. For the system with intermediate coupling, the higher energy discrete states with $|E_\alpha|<\Delta_0(\Gamma)$ (orange region) are represented by the dark-red lines, while the continuum above the bulk gap, $|E_\alpha|>\Delta_0(\Gamma)$, corresponds to the gray area. The results are obtained using a single chain model ($N_y=1$) with fixed chemical potential $\mu=0$ (relative to the bottom of the band).}
\vspace{-2mm}
\label{Fig05}
\end{figure}
%%%%%%%%%%%%%%%%%%%%%%	

The first significant effect of the proximity-induced low-energy renormalization is the suppression of the Majorana splitting oscillations, which is illustrated in Fig. \ref{Fig05}. We compare the oscillations that characterize a weak-coupling system with $\Delta_0 = 0.82$ meV and $\gamma=0.25 \Delta_0$ (dashed black lines) with the oscillations corresponding to an intermediate coupling regime with $\Delta_0 = 0.25$ meV and $\gamma=1.5 \Delta_0$ (red lines). Here, for simplicity, we use the notation $\Delta_0 = \Delta_0(0)$. The upper and lower panels  in Fig. \ref{Fig05} correspond to constant and field-dependent bulk SC gaps, respectively. Note that the amplitude of the oscillations increases with $\Gamma$ if $\Delta_0$ is constant, while for a field-dependent SC gap this trend reverses at large values of the Zeeman field.  

As a general observation, we note that one can rigorously define distinct phases only in the thermodynamic limit. By contrast, in finite systems the topological quantum phase transition (TQPT) is replaced by a crossover between a regime characterized by a large gap with no low-energy midgap states (the ``trivial phase'') and one characterized by the presence of a low-energy midgap  mode (the ``topological phase''). In Fig. \ref{Fig05} this crossover is marked by the lowest-field zero of the Majorana mode. As expected, the weak-coupling regime (dashed black line)  is characterized by a lower value of the crossover field as compared to the intermediate-coupling case (red line), consistent with the previous section, in particular Eq. (\ref{Gmmc}). Concerning higher energy states, we emphasize that the spectrum of a finite wire coupled to a hard-gap superconductor is always {\em discrete} below $\Delta_0(\Gamma)$. The ``trivial'' to ``topological'' crossover is signaled by a minimum of the second-lowest state. 
These properties are illustrated in Fig. \ref{Fig05} for the case $\gamma=1.5\Delta_0$. The field-dependence of the higher energy states is represented by the darker-red lines occupying the highlighted orange region. By contrast, above $\Delta_0(\Gamma)$ the spectrum is continuous (gray area). However, if the bulk superconductor has a finite density of sub-gap states, e.g., at finite magnetic field in the presence of disorder, the spectrum of the finite wire becomes continuous at all energies. If the density of sub-gap states is not too high, this would correspond to broadening the lines shown in Fig. \ref{Fig05}. These effects, which have crucial consequences from the perspective of quantum  computation, reveal the importance of understanding in detail the bulk superconductor and being able to optimize its properties. The bulk superconductor is most certainly not inert and benign!

Before we continue the discussion on the low-energy spectrum, it is instructive to clearly understand the significance of the  spectral lines. The energy $E_\alpha(\Gamma)$ of a low-energy state $\alpha$ is obtained by solving the equation ${\rm det}[G_{\rm eff}]=0$, i.e., finding the poles of the Green function. The corresponding wave function $\psi_\alpha$ is partially contained within the SM wire, but it also extends into the bulk SC. The total spectral weight contained inside the wire is
\begin{equation}
Z_\alpha \equiv \sum_{{\bm i}\in {\cal L}_{\rm SM}} |\psi_\alpha({\bm i})|^2= \left(1+\frac{\gamma}{\sqrt{\Delta_0^2-E_\alpha^2}}\right)^{-1}.
\end{equation}
In the weak-coupling limit, $\gamma\ll \Delta_0$, we have $Z_\alpha\rightarrow 1$, i.e., the state resides (almost) entirely inside the SM wire. By contrast, in the strong coupling limit, $\gamma \gg \Delta_0$, most of the spectral weight is inside the bulk SC. For example, a Majorana mode ($E_M = 0$) will only have a fraction $Z_M = \Delta_0/(\Delta_0+\gamma) \approx \Delta_0/\gamma$ of the wave function inside the wire. 

The information about the partition of the wave function between the wire and the bulk SC can be obtained by calculating the density of states inside the wire,
\begin{equation}
\rho(\omega)\equiv -\frac{1}{\pi}{\rm Im}\left\{{\rm Tr}[G_{\rm eff}(\omega+i\eta)]\right\} = \sum_\alpha Z_\alpha \delta(\omega-E_\alpha),
\end{equation}
where $\eta\rightarrow 0^+$ and $|E_\alpha|<\Delta_0(\Gamma)$. In the weak-coupling regime the weight of the poles is close to one, while for intermediate/strong coupling the weight is significantly reduced, i.e., the quasiparticle wave functions leak into the SC. Note that above $\Delta_0(\Gamma)$ there is a continuum of states, each of them residing almost entirely inside the bulk SC, but having a small tail inside the SM wire.

%%%%%%%%%%%%%%%%%%%%%%%%%%%%
\begin{figure}[t]
\begin{center}
\includegraphics[width=0.48\textwidth]{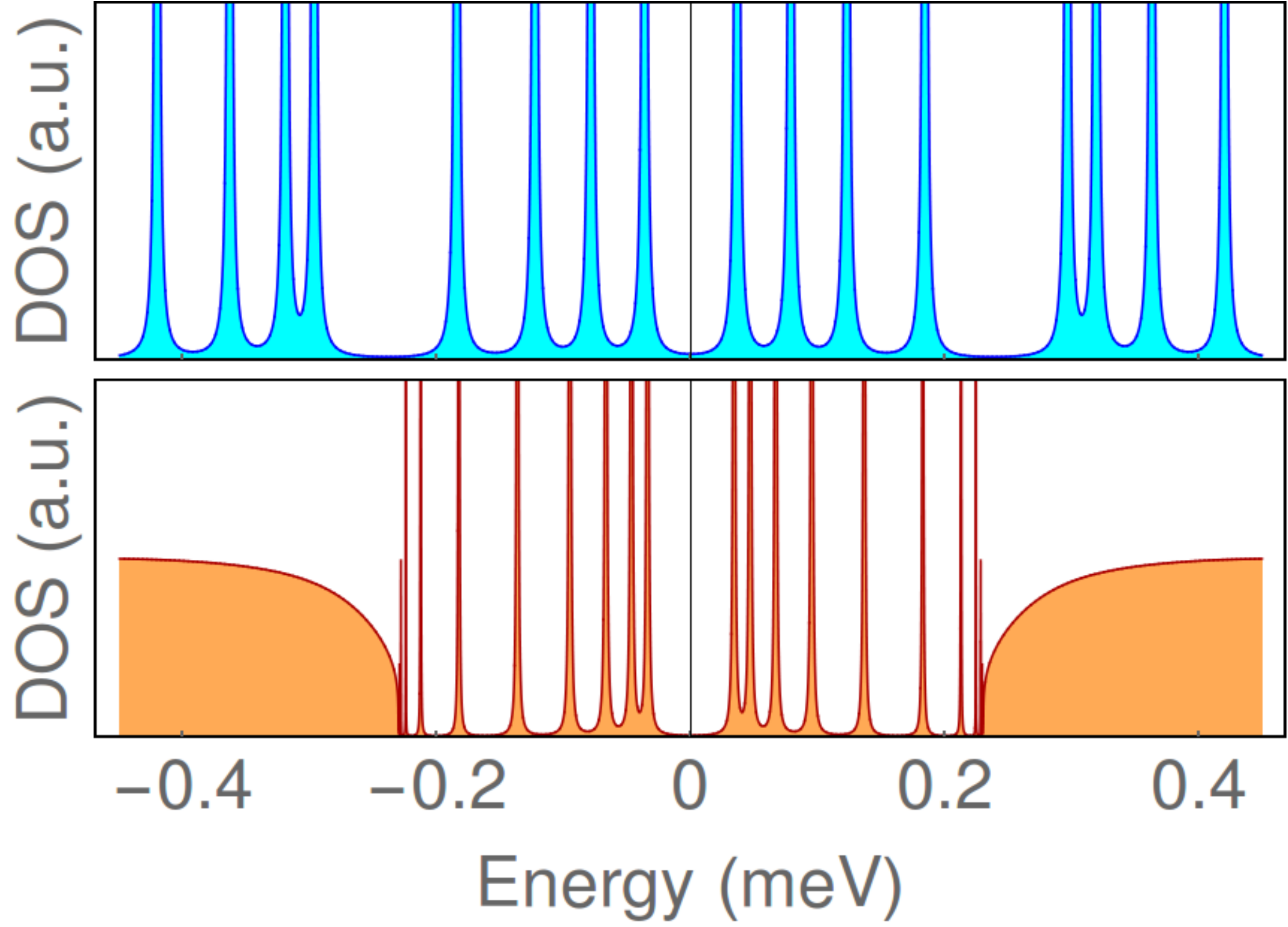}
\vspace{-3mm}
\end{center}
\caption{(Color online) Density of states of the SM wire proximity-coupled to a superconductor. The wire is modeled as a single chain ($N_y=1$). {\em Top}: Weak coupling regime with $\gamma=0.167$ meV and $\Delta_0 \rightarrow \infty$. {\em Bottom}: Intermediate coupling with $\gamma = 1.5\Delta_0$, $\Delta_0 = 0.25$ meV. A finite  Zeeman field $\Gamma=0.8\gamma_c$ is applied. The delta peaks were broaden by hand to reveal their weight.}
\vspace{-2mm}
\label{Fig06}
\end{figure}
%%%%%%%%%%%%%%%%%%%%%%	

A specific example is shown in Fig. \ref{Fig06}. The two panels show the density of states of a SM wire coupled to a SC and in the presence of a finite Zeeman  field. In the bottom panel the SC gap is $\Delta_0=0.25$ mev and the SM-SC coupling is $\gamma=1.25 \Delta_0$, while the top panel corresponds to the weak coupling regime  with $\gamma = 0.167$ meV and $\Delta_0\rightarrow \infty$. Note that these parameters correspond to the same value of the induced gap, $\Delta_{\rm ind}=0.167$ meV. Three important features deserve to be mentioned. First, the  spectrum of the SM wire is discrete below the bulk SC gap $\Delta_0$ and continuous above, as evident from the lower panel. Second, the energy difference between the low-energy states decreases with increasing the ratio $\gamma/\Delta_0$. Within the energy interval $(-0.25, 0.25)$ meV there are roughly twice as many states in the bottom panel as compared to the top panel. This is the most direct manifestation of the proximity-induced low-energy renormalization and a consequence of the fact that proximity-coupling modifies the frequency term in the Green function denominator according to $\omega \longrightarrow \omega(1+\gamma/\sqrt{\Delta_0^2-\omega^2})$. Third, the spectral weight $Z_\alpha$ inside the wire is reduced by the SM-SC coupling, as one can easily see by comparing the width of the artificially-broadened peaks from top and bottom panels in Fig. \ref{Fig06}. Of course, spectra, like those in Fig. \ref{Fig05}, do not contain information regarding the reduced spectral weight of various states, but this feature may be important when discussing the ``visibility'' of certain modes, e.g., when calculating tunneling conductances and local densities of states. 

%%%%%%%%%%%%%%%%%%%%%%%%%%%%
\begin{figure}[t]
\begin{center}
\includegraphics[width=0.48\textwidth]{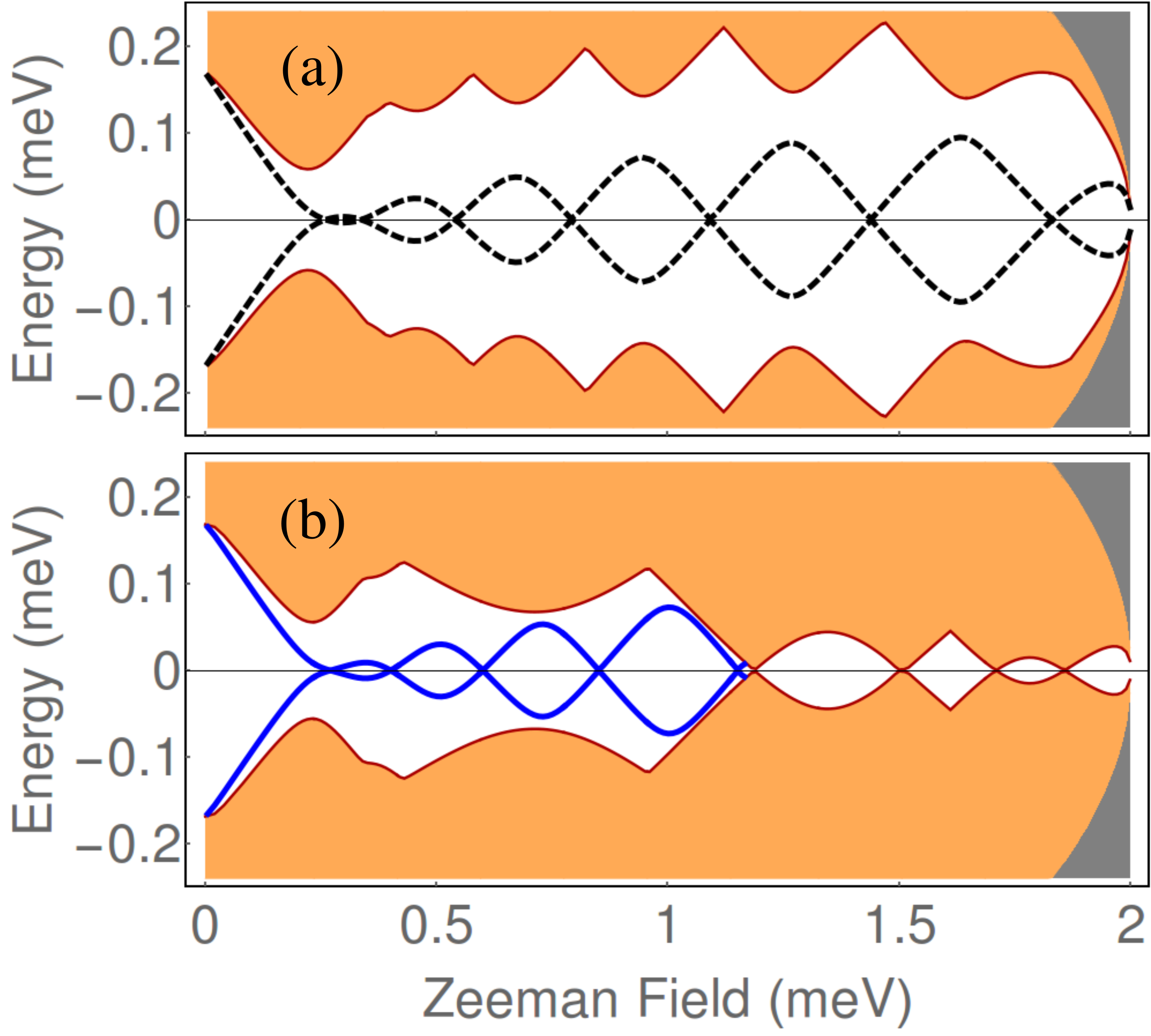}
\vspace{-5mm}
\end{center}
\caption{(Color online) Dependence of the lowest energy mode and first excited state on the Zeeman field in the weak-coupling regime with $\Delta_0=0.82$ meV and $\gamma=0.25\Delta_0$. The SC gap in field dependent and collapses at $\Gamma = 2$ meV (gray area). (a) Single band model and chemical potential at the bottom of the band. (b) Two-band model with the chemical potential at the bottom of the second band. The dashed black line is the same as in the bottom panel of Fig. \ref{Fig05}.  The collapse of the quasiparticle gap in (b) is due to states from the lower energy occupied band. In the orange regions there are higher discrete states that are not shown explicitly.}
\vspace{-3mm}
\label{Fig07}
\end{figure}
%%%%%%%%%%%%%%%%%%%%%%	

Returning to the Zeeman field dependence of the low-energy spectrum, a natural question concerns the possible differences between the single band case shown in Fig. \ref{Fig05} and a multi-band system. To address this question, we first consider a two-chain model ($N_y=2$) in the independent band regime with uniform coupling $\bar{\gamma}_{\bm i} = \gamma$. The comparison with the single band case is shown in Fig. \ref{Fig07}. Two important features deserve some comments. First, we note that the lowest energy (Majorana) mode has practically the same dependence on the Zeeman field in the top and bottom panels, before the collapse of the quasiparticle gap in (b). This is due to the fact that in the independent band regime Majorana physics is controlled by the top occupied band and only depends on the position of the chemical potential relative to that band. Second, we note that the quasiparticle gap separating the Majorana mode from higher energy quasiparticles is significantly suppressed in the two-band model. This suppression can be minimized by increasing the strength of the Rashba spin-orbit coupling (here $\alpha_R = 150$ meV$\cdot$\AA) and the inter-band separation (here $\Delta\epsilon=4$ meV). In the limit $\Delta\epsilon\rightarrow \infty$ we recover the single band result. In practice, a real concern  should be the presence of an occupied band that is very close in energy to the top (Majorana) band.  The presence of such a band would compromise the quasiparticle gap that protects the Majorana mode and thus the possibility of topologically-protected operations. While the lowest energy nanowire band is well separated from the higher energy spectrum, in wires with multi-band occupancy the occurrence of small inter-band gaps in the vicinity of the chemical potential is generically expected.

%%%%%%%%%%%%%%%%%%%%%%%%%%%%
\begin{figure}[t]
\begin{center}
\includegraphics[width=0.48\textwidth]{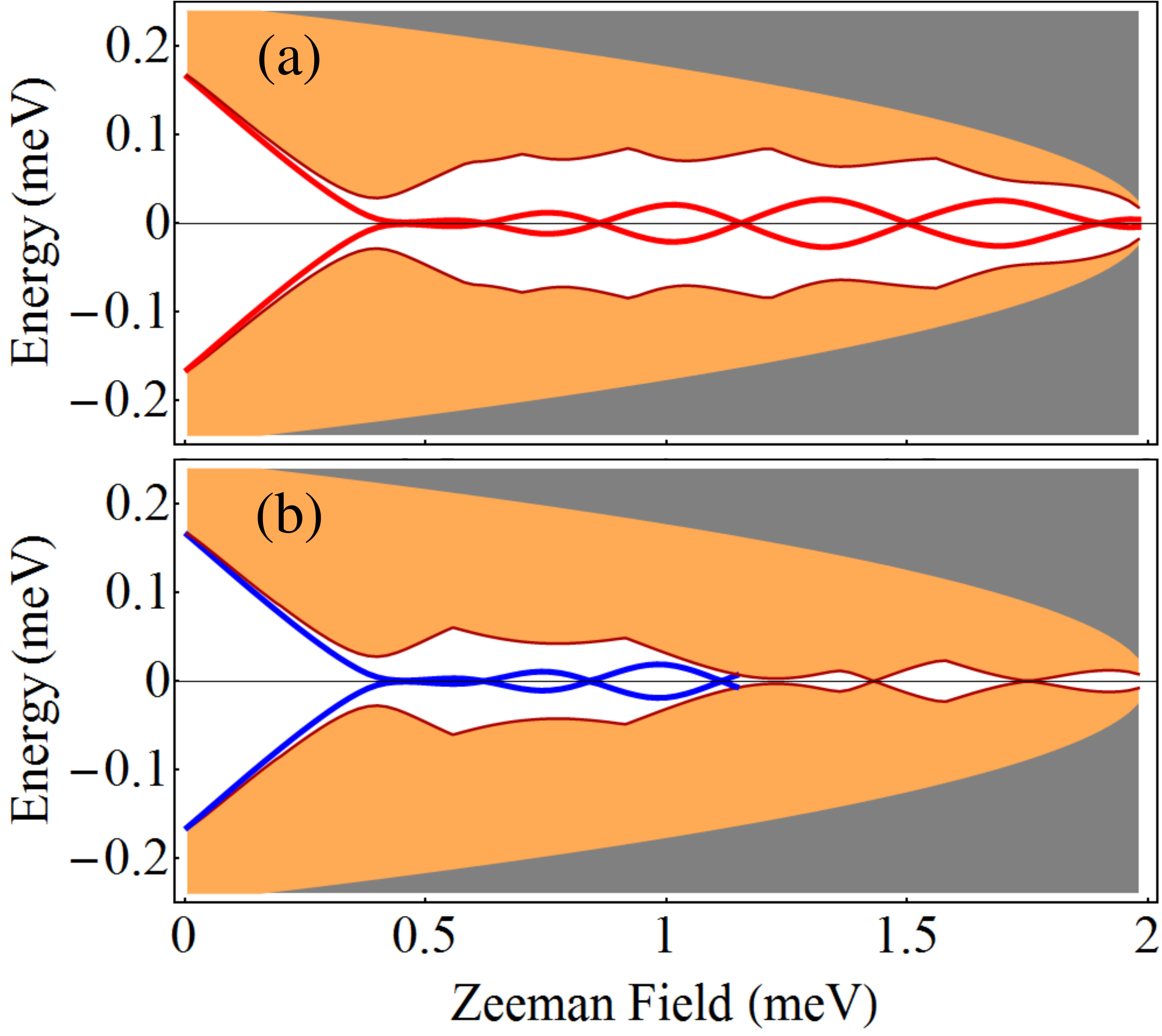}
\vspace{-5mm}
\end{center}
\caption{(Color online) Zeeman field dependence of the lowest two energy modes at intermediate coupling ($\gamma=1.5\Delta_0$, with $\Delta_0=0.25$ meV) for a single-band system (a) and a two-band system in the independent band regime (b). The top panel shows the same results as the lower panel of Fig. \ref{Fig05} (note the different energy scale).}
\vspace{-4mm}
\label{Fig08}
\end{figure}
%%%%%%%%%%%%%%%%%%%%%%	

Next, we consider a two-band system with intermediate coupling, $\gamma=1.5\Delta_0$, where $\Delta_0=0.25$ meV. The comparison with the single band case is shown in Fig. \ref{Fig08}. The main  features are qualitatively the same as in Fig. \ref{Fig07}, i.e., the Majorana modes in single- and multi-band systems have similar properties (at low-enough values of the Zeeman field) and the quasiparticle gap is reduced in the multi-band case. Quantitatively, however, there is a major difference between the weak coupling regime shown in Fig. \ref{Fig07} and the intermediate-coupling case illustrated in Fig. \ref{Fig08}. The Majorana splitting oscillations and the magnitude of the quasiparticle gap are reduced more than twice in the intermediate coupling case (Fig. \ref{Fig08}) as compared to weak coupling (Fig. \ref{Fig07}). This significant difference occurs as a result of the low-energy proximity-induced renormalization, despite the fact that the hybrid systems are characterized by the same value of the induced gap $\Delta_{\rm ind}=0.167$ meV. Indeed, at zero magnetic field the energy of the lowest-energy state (i.e., the induced gap) is the same for all four systems shown in Fig. \ref{Fig07} and Fig. \ref{Fig08}. 

%%%%%%%%%%%%%%%%%%%%%%%%%%%%
\begin{figure}[t]
\begin{center}
\includegraphics[width=0.48\textwidth]{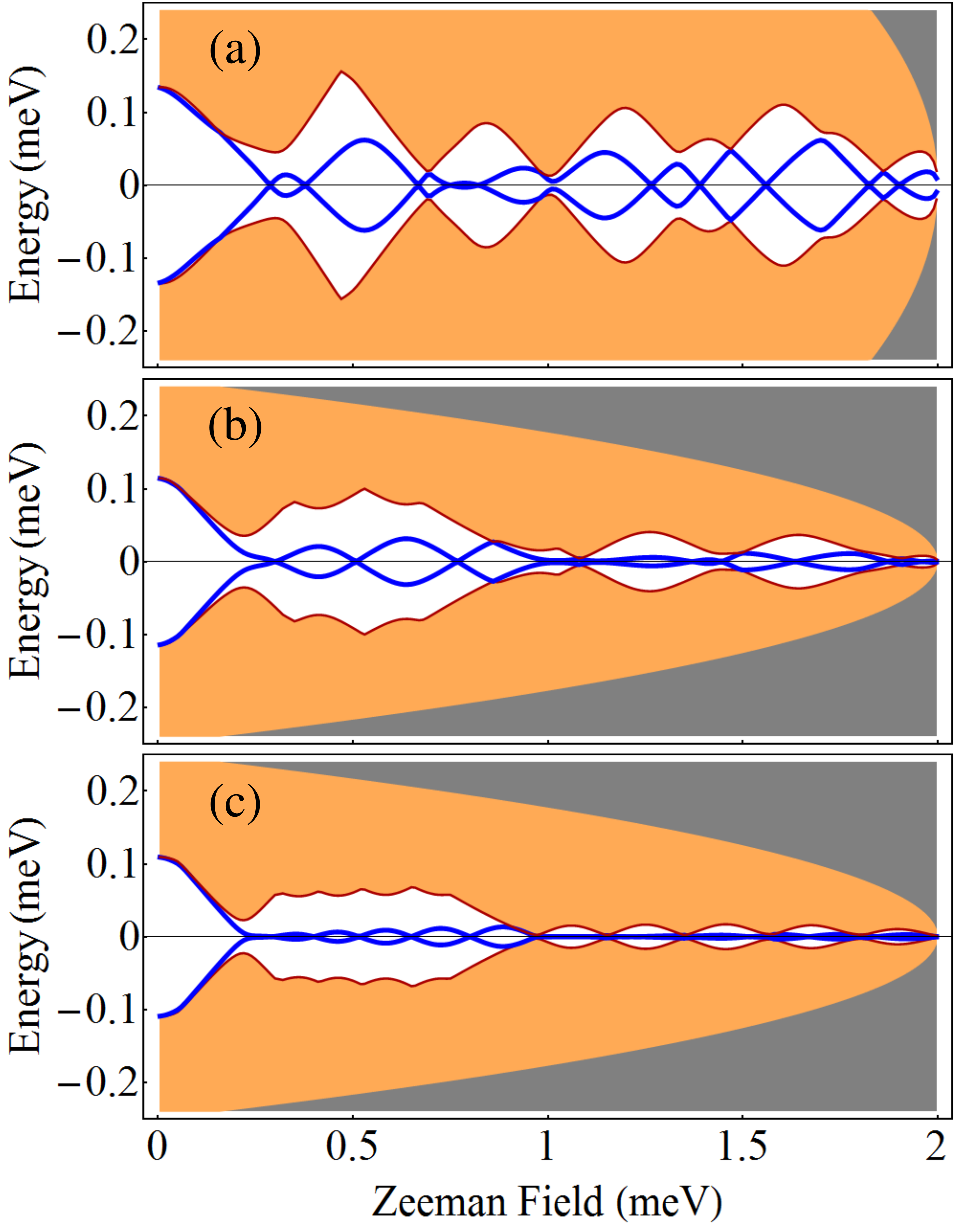}
\vspace{-5mm}
\end{center}
\caption{(Color online) Low-energy modes in the coupled band regime described by Eq. (\ref{gma}) for a system with inter-band spacing $\Delta\epsilon=0.5$  meV.  (a) Weak coupling: $\Delta_0 = 0.82$ meV, $\gamma=0.25\Delta_0=0.205$ meV. (b) Intermediate coupling: $\Delta_0=0.25$ meV, $\gamma=1.5\Delta_0=0.375$ meV. (c) Same as (b) but for a longer wire, $L_x=1.5$ $\mu$m. The chemical potential is tuned midway between the two bands.}
\vspace{-4mm}
\label{Fig09}
\end{figure}
%%%%%%%%%%%%%%%%%%%%%%	

The final scenario that we investigate here involves a two-band system in the coupled band regime. This scenario is potentially relevant in multi-band systems whenever the chemical potential is in the vicinity of nearly-degenerate bands. The Zeeman field dependence of the lowest two energy states is shown in Fig. \ref{Fig09} for three different sets of parameters. The first striking feature is that the crossover Zeeman field $\Gamma_{c1}$ decreases when increasing the coupling strength $\gamma$, in contrast with the generic behavior that characterizes the independent band regime. This behavior is consistent with the conclusions of Sec. \ref{IW} regarding the infinite system. Note that panels (b) and (c) in Fig. \ref{Fig09} correspond to vertical cuts at $\mu=0$ in the phase diagram shown in Fig. \ref{Fig03} (c) for finite-size systems with $L_x=0.8$ $\mu$m and   $L_x=1.5$ $\mu$m, respectively. The transverse spin-orbit coupling $\alpha_R^\prime=0.75$ meV gives a lower critical field corresponding to the red line in Fig. \ref{Fig03} (c), i.e. $\Gamma_{c1}\approx 0.21$ meV, while the corresponding crossover fields are slightly higher due to finite size effects. 

The emergence of  weakly- and  strongly-coupled ``bands''  as a result of the proximity-induced inter-band coupling (Sec. \ref{IW})  is nicely illustrated in Fig. \ref{Fig09} (b) and (c). The weakly coupled band is responsible for the lower value of the crossover field $\Gamma_{c1}$ (slightly larger than $0.2$ meV) and produces a Majorana mode with sizable splitting oscillations (at least in the short wire). The strongly-coupled band controls $\Gamma_{c2}$ (around $1$ meV)  and generates a second Majorana mode with significantly lower amplitude oscillations. Focusing on the weakly coupled band, i.e., Zeeman fields $\Gamma < 1$ meV, we note that the corresponding Majorana splitting oscillations are larger than the oscillations characterizing a system in the independent band regime having the same nominal SM-SC coupling, $\gamma=1.5\Delta_0 = 0.375$ meV, as evident when comparing Fig. \ref{Fig09} (b) and  Fig. \ref{Fig08} (b). The quasiparticle gap is also  larger in the coupled band regime. On the other hand, the splitting oscillations in  Fig. \ref{Fig09} (b) have a lower amplitude than the oscillations characterizing a nominally weakly coupled system with a comparable crossover field $\Gamma_{c1}$ (see Fig. \ref{Fig07}). 

%%%%%%%%%%%%%%%%%%%%%%%%%%%%
\begin{figure}[t]
\begin{center}
\includegraphics[width=0.48\textwidth]{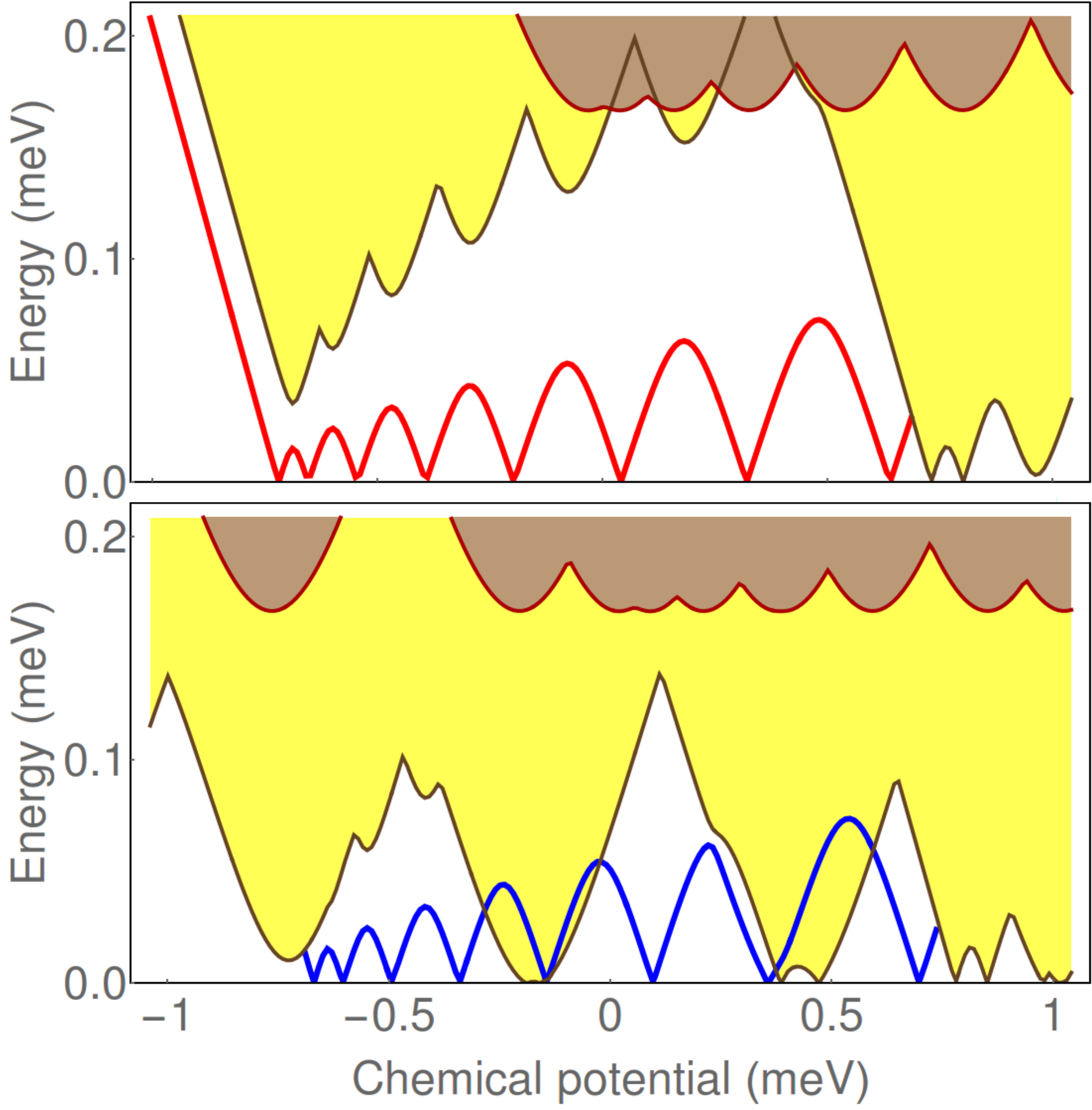}
\vspace{-5mm}
\end{center}
\caption{(Color online) Dependence of the low-energy states on the chemical potential for a system with weak SM-SC coupling, $\gamma = 0.25\Delta_0$, with $\Delta_0=0.82$ meV. The top panel corresponds to a single-band system ($N_y=1$), while the bottom panel is for a system with $N_y=2$ in the independent band regime. The boundary of the brown region represents the lowest energy state at zero magnetic field and its minimum corresponds to an induced gap $\Delta_{\rm ind}\approx 0.167$ meV. The red/blue line represents the Majorana mode at $\Gamma=0.75$ meV and the boundary of the yellow region  is the first excited state at the same Zeeman field. The chemical potential is measured relative to the bottom of the band (second band for the bottom panel).}
\vspace{-3mm}
\label{Fig10}
\end{figure}
%%%%%%%%%%%%%%%%%%%%%%	

\subsection{Dependence on the chemical potential and ``phase diagram''}

In the previous section we have investigated the low-energy physics of a proximity-coupled wire by varying the applied Zeeman field while maintaining a constant chemical potential. We have shown that the Majorana splitting oscillations and the quasiparticle gap are significantly renormalized when the effective SM-SC coupling becomes comparable with (or larger than) the bulk SC gap, i.e., in the intermediate/strong coupling regime. For completeness, we consider now the dependence of the low-energy features on the chemical potential at fixed values of the Zeeman field. More specifically, we calculate the dependence of the lowest-energy states on the chemical potential in the weak and intermediate coupling regimes for single-band and two-band systems (within the independent band regime) at zero magnetic field and in the presence of a finite Zeeman field $\Gamma=0.75$ meV. 

\begin{figure}[t]
\begin{center}
\includegraphics[width=0.48\textwidth]{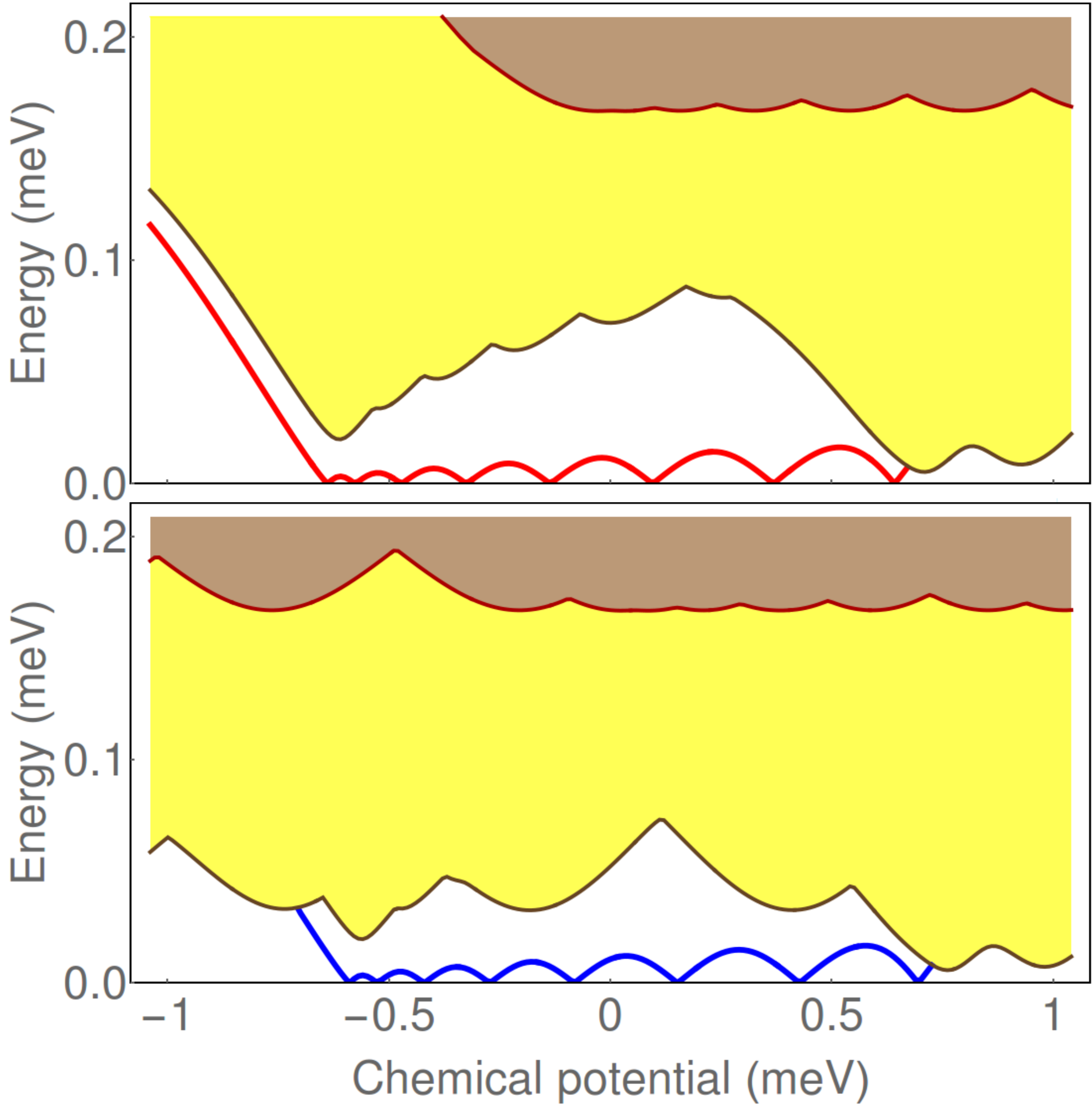}
\vspace{-5mm}
\end{center}
\caption{(Color online) Dependence of the low-energy states on the chemical potential for a system with intermediate SM-SC coupling, $\gamma = 1.5\Delta_0$, with $\Delta_0=0.25$ meV. The top and bottom panels correspond to $N_y=1$ and $N_y=2$, respectively. The significance of the symbols is the same as in Fig. \ref{Fig10}.}
\vspace{-2mm}
\label{Fig11}
\end{figure}

The weak coupling behavior is illustrated in Fig. \ref{Fig10}, while the intermediate coupling results are shown in Fig. \ref{Fig11}. First, we note that induced gap, i.e. the minimum of the lowest-energy state at $\Gamma=0$, is the same in all four panels, in agreement with our analysis of the infinite system in Sec. \ref{IW}. Note that the boundary of the brown regions from the top panels (i.e., the lowest state of a single-band system at $\Gamma=0$) has the same structure as induced gap shown in Fig. \ref{Fig02}. The upturn associated with the depletion of the band has a slope that is renormalized by the SM-SC coupling. The weak dependence on $\mu$ for positive values of the chemical potential is a finite size effect reflecting the discreteness of the quantum states in a (relatively) short wire.   The additional  features present in the bottom panels are generated by states from the lower energy occupied band. 

%%%%%%%%%%%%%%%%%%%%%%%%%%%%
%%%%%%%%%%%%%%%%%%%%%%

The major differences between the weak and the intermediate coupling regimes emerge in the presence of a finite Zeeman field. First, we note that the amplitude of the Majorana splitting oscillations is significantly suppressed at intermediate coupling, $\gamma=1.5\Delta_0$, as compared to weak coupling, $\gamma=0.25\Delta_0$. A direct consequence  of this property is that the experimental observation of these oscillations in strongly-coupled hybrid systems may be difficult, as it requires high energy resolution. Note that, indeed, experimental observations of Majorana oscillations are rare.
A second consequence of increasing the SM-SC coupling is the reduction of the quasiparticle gap that protects the Majorana mode. In the example shown in Fig. \ref{Fig11}, this quasiparticle gap is of the order of $40-90$ $\mu$eV. We emphasize that these values correspond to an ``ideal'', i.e., perfectly clean, system. Taking disorder into account will generate additional low-energy states, thus further reducing the magnitude of the quasiparticle gap. In turn, a small quasiparticle gap represents a serious challenge for implementing topologically-protected quantum operations using Majorana zero modes. These observations suggest that using a low-gap superconductor (e.g., aluminum) strongly coupled to the semiconductor wire may not be an optimal path toward engineering a topological qubit, despite the fact that it generates a sizable induced gap at zero magnetic field. At finite field, strong coupling will inevitably bring in sub-gap states from the bulk superconductor, considerably complicating the low-energy SM spectrum and destoying the topological protection of the Majorana modes.

The main results regarding the proximity-induced low-energy renormalization can be summarized as an effective ``phase diagram'' showing the  
 dependence of the lowest-energy state of a finite hybrid system on the Zeeman field and chemical potential.  In a long-enough wire, the transition to a topological phase at the lower critical field $\Gamma_{c1}$ is signaled by the vanishing of the quasiparticle gap associated with the emergence of a zero-energy Majorana mode. In the presence of a non-vanishing transverse spin-orbit coupling, $\alpha_R^\prime\neq 0$, higher field phase boundaries (e.g., $\Gamma_{c2}$) are associated with the opening/closing of the small ``minigap''  that characterizes a wire with an even number of low-energy (Majorana) modes at each end. In practice, observing these higher field transitions is difficult due to finite size effects (in small systems), energy resolution requirements, or large values of the Zeeman field, which could destroy bulk superconductivity itself.  Also note that regions characterized by small values of the quasiparticle gap may occur inside the low-field ``trivial phase'' (see below), hence extra care is always required when interpreting experimentally-measured ``phase diagrams'' \cite{Chen2016}. 	
 
%%%%%%%%%%%%%%%%%%%%%%%%%%%%
\begin{figure}[t]
\begin{center}
\includegraphics[width=0.47\textwidth]{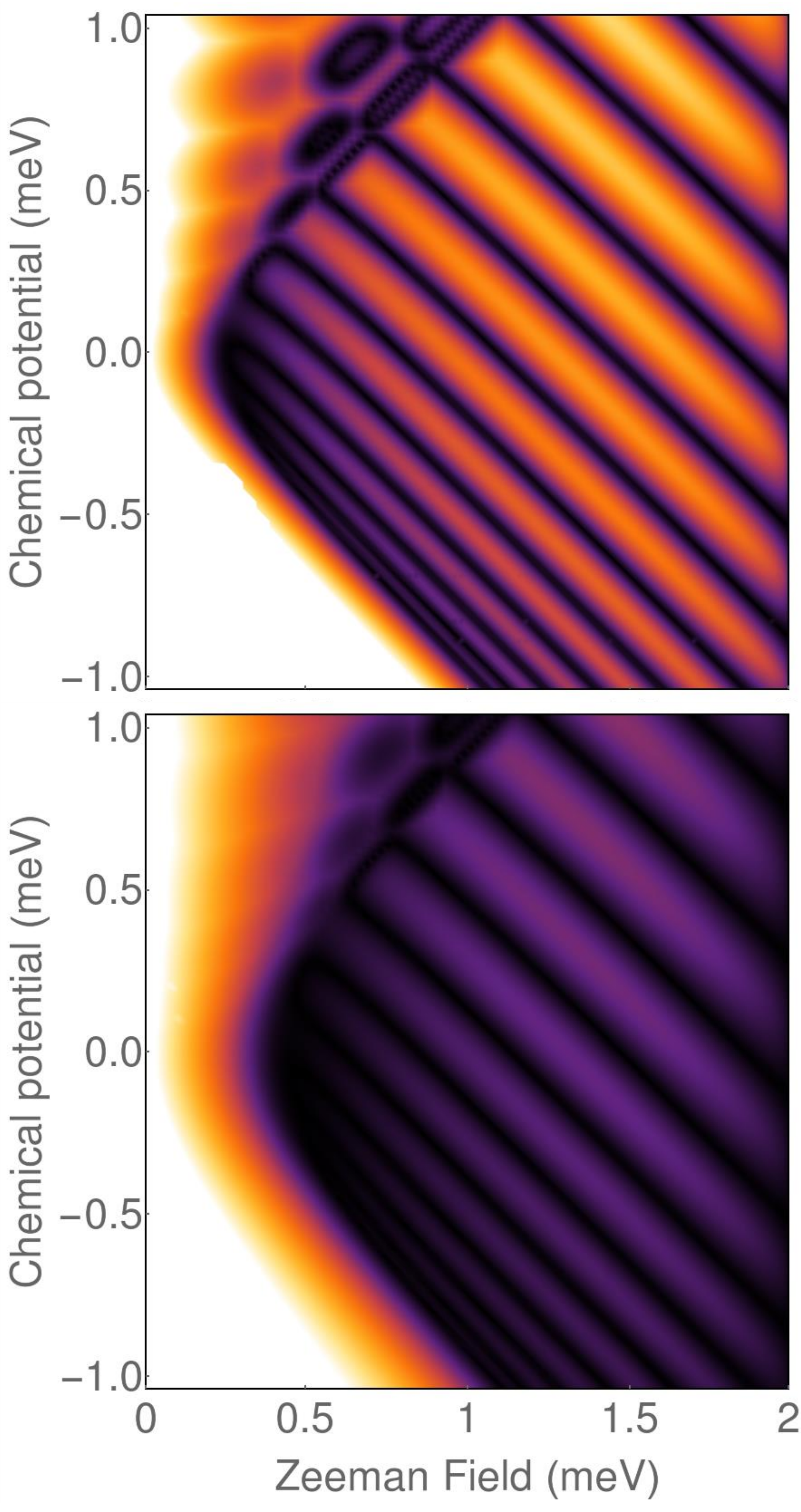}
\vspace{-5mm}
\end{center}
\caption{(Color online) Effective phase diagram of a finite system as revealed by the dependence of the lowest energy state on Zeeman field and chemical potential. {\em Top}: Weak coupling regime ($\gamma=0.25\Delta_0$, $\Delta_0=0.82$ meV). {\em Bottom}: Intermediate coupling regime ($\gamma=1.5\Delta_0$, $\Delta_0=0.25$ meV). White corresponds to energies larger than $0.1$ meV, while black gapless states.
The ``stripy'' region, which corresponds to the ``topological phase'', reveals the Majorana splitting oscillations. Note that the oscillations are significantly suppressed at intermediate coupling (bottom) as compared to weak coupling (top). In addition, the ``phase boundary'' changes with $\gamma$, consistent with Eq. (\ref{Gmmc}). The color code is the same in both panels; white corresponds to energies $E\geq 0.15$ meV.}
\vspace{-2mm}
\label{Fig12}
\end{figure}
%%%%%%%%%%%%%%%%%%%%%% 				

A comparison between the ``phase diagrams'' in the weak- and intermediate-coupling regimes is shown in Fig. \ref{Fig12}. The results are for simgle-band systems with $\gamma=0.25\Delta_0$ (top) and $\gamma=1.5\Delta_0$ (bottom). For a vanishing Zeeman field, the energy of the lowest-energy state is equal to or larger than the induced gap $\Delta_{\rm ind}\approx 0.17$ meV. This energy decreases with increasing Zeeman field and  vanishes at a the crossover field $\Gamma_{c1}(\mu)$, which is slightly larger than the value predicted by Eq. (\ref{Gmmc}) due to finite size effects. At higher values of the applied Zeeman field the system is in the ``topological phase''  and the lowest energy state corresponds to a pair of weakly overlapping Majorana modes localized at the ends of the wire. The corresponding energy splitting oscillations  generate the ``stripy'' regions in Fig. \ref{Fig12}.

%%%%%%%%%%%%%%%%%%%%%%%%%%%%
\begin{figure}[t]
\begin{center}
\includegraphics[width=0.48\textwidth]{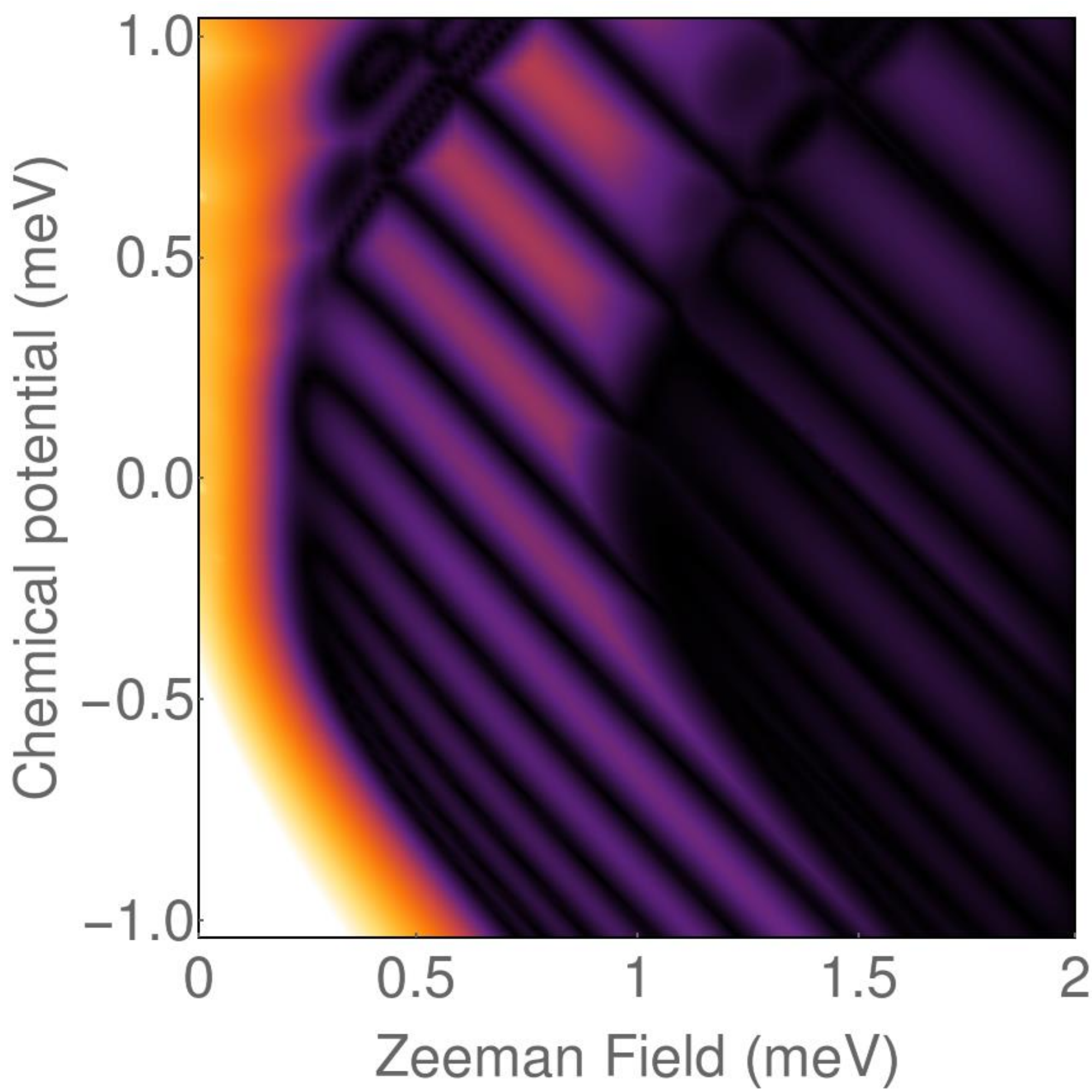}
\vspace{-5mm}
\end{center}
\caption{(Color online) Effective phase diagram of a finite system in the coupled band regime described by Eq. (\ref{gma}) with $\Delta\epsilon=0.5$ meV and $\gamma=1.5\Delta_0=0.375$ meV. The transverse Rashba coefficient is $\alpha_R^\prime=0.75$ meV, so that the lower crossover field $\Gamma_{c1}$ corresponds to the red line from Fig.\ref{Fig03} (c).}
\vspace{-2mm}
\label{Fig13}
\end{figure}
%%%%%%%%%%%%%%%%%%%%%%
	
A few remarks are warranted. First, note the striking difference between the amplitudes of the Majorana splitting oscillations at weak- and intermediate-coupling (top and bottom panels of Fig. \ref{Fig12}, respectively), which is a direct consequence of the proximity-induced energy renormalization. We emphasize that in the absence of a Zeeman field the  lowest energy  has similar values in both cases. Second, we note that the boundaries of the ``stripy'' regions are consistent with the topological phase boundary described by Eq. (\ref{Gmmc}). Third, the region of the ``trivial phase'' corresponding to $\mu, \Gamma >0.4$ meV is characterized by the presence of a low-energy state (dark spots outside the ``stripy'' areas  in Fig. \ref{Fig12}). This state is associated with the higher energy spin sub-band, which approaches zero-energy in the vicinity of the trivial-to-topological crossover, as revealed by the behavior of the yellow regions in Figs. \ref{Fig10} and \ref{Fig11} (top panels). Finally, we note that the Majorana splitting oscillations should be observable provided i) their amplitude is larger than the experimental energy resolution, which can always be realized by working with short-enough wires in the weak coupling regime, and ii) the measurement is done along a path in the phase diagram that crosses the ``stripes''. Concerning the second requirement, we note that varying the Zeeman field does not guarantee a constant chemical potential due to electrostatic effects.  This may result in paths that run parallel to the stripes, which conceals the presence of the energy splitting oscillations (see next section). On the other hand, the chemical potential can be varied while maintaining a constant value of the Zeeman field. This corresponds to vertical cuts of in the ``phase diagram'', which necessarily leads to splitting oscillations (see, for example, Figs. \ref{Fig10} and \ref{Fig11}).  

A ``phase diagram'' corresponding to a two-band system in the coupled band regime is shown in Fig. \ref{Fig13}. 	
Its structure can be understood by combining the information about the phase diagram of the infinite system shown in Fig. \ref{Fig03} (c) with the field dependence of the low-energy spectrum, which is shown in Fig. \ref{Fig09} (b). The ``topological phase'' corresponds to the ``stripy'' area, while the dark high-field region represents a trivial ``phase'' characterized by the presence of two Majorana modes at each end of the wire. Note that for chemical potentials within the interval $(-0.5, 0.5)$ meV the (low-field) trivial-to-topological crossover takes place at low  values of the Zeeman field and $\Gamma_{c1}$  is weakly dependent on the chemical potential. This behavior is strikingly different from that of phase boundaries in the independent band regime (see Fig. \ref{Fig12}). Finally, we note that, despite the low values of the crossover field, the amplitude of the Majorana oscillations is significantly lower than the corresponding amplitude in the weakly coupled independent-band regime, as evident when comparing Fig. \ref{Fig13} and the top panel of Fig. \ref{Fig12} (the same color code was used in both figures).

\subsection{The ``visibility'' of low-energy features}
																						
The last topic that we address concerns the experimentally-observable consequences on the low-energy features discussed in the preceding sections. More specifically, we focus on charge tunneling experiments, which, basically, probe the local density of states at the end of the proximitized SM wire. There are two key ingredients that play critical roles in determining the tunneling conductance and the local density of states (LDOS), ingredients that have not been considered so far in our discussion.   The first one concerns the spatial distribution of the spectral weight along the wire. Of course, energy spectra, like those in Figs. \ref{Fig07}-\ref{Fig11}, do not contain this type of information. However, it is straightforward to extract it from the effective Green function by calculating the LDOS
\begin{equation}
\rho(\omega, {\bm i}) = -\frac{1}{\pi}{\rm Im}\left\{{\rm Tr}[G_{\rm eff}(\omega+i\eta, {\bm i}, {\bm i})] \right\},
\end{equation}																						where ${\rm Im}\{\dots\}$ and 	${\rm Tr}[\dots]$ represent the imaginary part and the trace over spin and particle-hole degrees of freedom, respectively. 

The second aspect concerns the description of the bulk superconductor. In this study we use the simplified model of the bulk SC given by Eq. (\ref{Sigma}), which does not include effects due to the presence of disorder and finite magnetic fields. However, in experimentally-relevant conditions  these effects may be crucial. In essence, the presence of disorder and finite magnetic field is expected to generate low-energy, sub-gap states in the bulk SC. Once the SC is coupled to the SM wire, these low-energy states hybridize with the states residing inside the wire, effectively resulting in a broadening of the spectral features at all energy scales, i.e., including energies lower than $\Delta_0(\Gamma)$. Appropriately modeling the bulk SC to incorporate the effects of disorder and finite magnetic field remains a crucial outstanding problem in this field \cite{Cole2016,Hui2015}. Here, we do not address this issue at the microscopic level. Instead, we use a toy model that describes the sub-gap states in terms of an imaginary contribution to the SC Green function that is obtained through the substitution $\omega \rightarrow \omega+ i \eta(\Gamma)$, where $\eta(\Gamma)$ is a field-depended energy scale related to the presence of low-energy SC states. In the numerical calculations, the dependence of this energy scale  on the Zeeman field is 
\begin{equation}
\eta(\Gamma) = 0.01+ 0.025 \Gamma^2/(0.2+\Gamma^2) ~~{\rm meV}, \label{eta}
\end{equation}
which increases with the applied field, ranging from $10$ $\mu$eV to $35$ $\mu$eV. This field dependence is chosen to simulate the expected increase of the density of low-energy states in the bulk SC with the applied magnetic field.

%%%%%%%%%%%%%%%%%%%%%%%%%%%%
\begin{figure}[t]
\begin{center}
\includegraphics[width=0.48\textwidth]{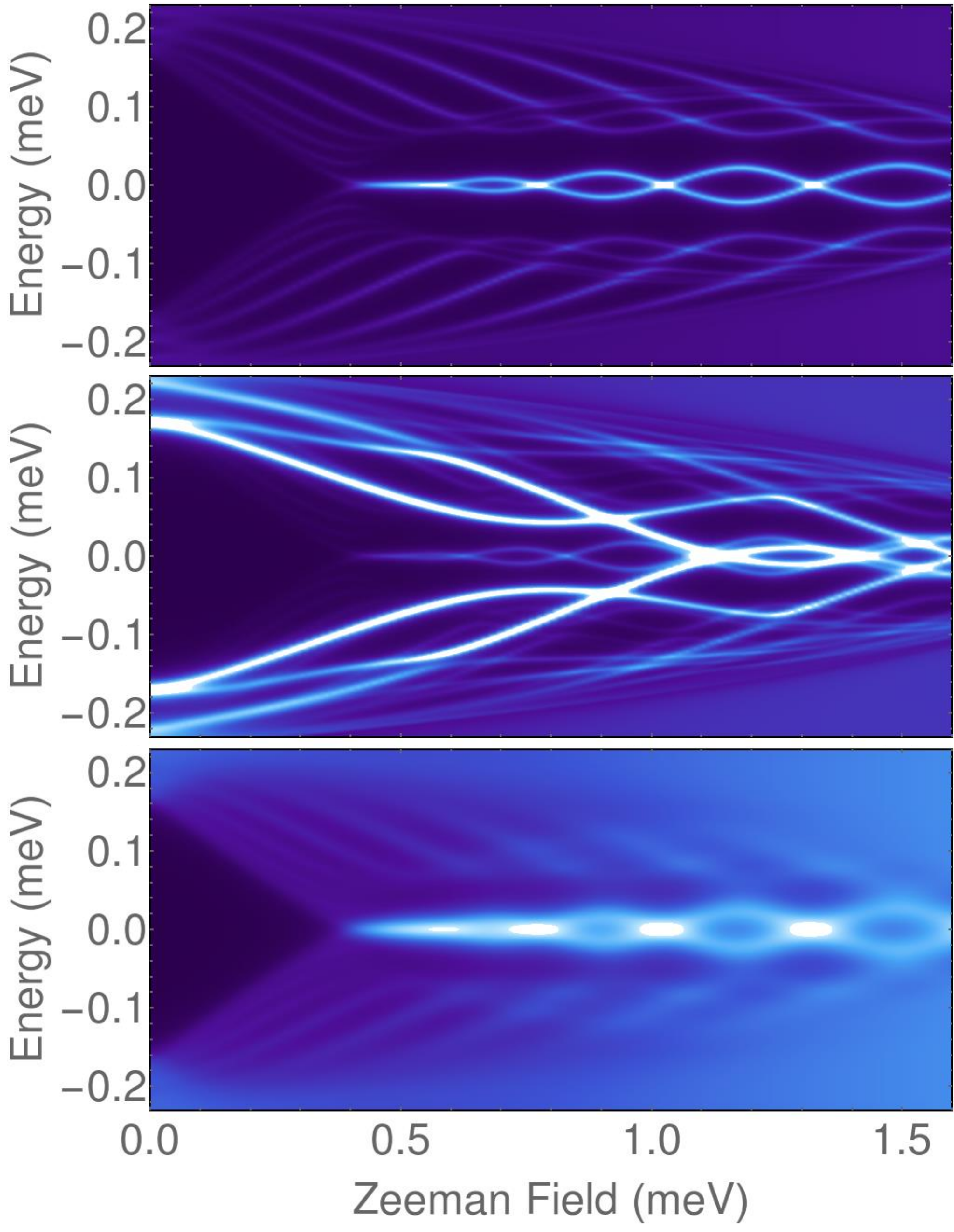}
\vspace{-5mm}
\end{center}
\caption{(Color online) Local density of states at the end of the wire as function of energy and Zeeman field. {\em Top}: Single band system with the same parameters as in Fig. \ref{Fig08} (a) and a ``clean'' superconductor with $\eta(\gamma) = 0.005$ meV. Note the higher visibility of the Majorana mode as compared to the ``bulk'' states. {\em Middle}: Two-band system with same parameters as in Fig. \ref{Fig08} (b) and  ``clean'' superconductor.  The states associated with the low-energy occupied band have higher visibility that states corresponding to the top (Majorana) band. {\em Bottom}: Single band wire coupled to a ``dirty'' superconductor. The model parameters are the same as in the top panel and $\eta(\Gamma)$ is given by Eq. (\ref{eta}).}
\vspace{-2mm}
\label{Fig14}
\end{figure}
%%%%%%%%%%%%%%%%%%%%%%
	
Consider now a single-band system in the intermediate-coupling regime corresponding to the parameters of Fig. \ref{Fig08} (a), with a  self-energy (\ref{Sigma}) that includes the substitution $\omega \rightarrow \omega+ i \eta(\Gamma)$,  with $\eta(\Gamma)$ given by Eq. (\ref{eta}). The dependence of the LDOS at the end of the wire on energy and Zeeman field for a hybrid system with a relatively ``clean'' superconductor characterized by $\eta(\Gamma) = 0.005$ meV is shown in the top panel of Fig. \ref{Fig14}. The model parameters are the same as in the bottom panel of Fig. \ref{Fig05} and in Fig. \ref{Fig08} (a). Indeed, upon close inspection one can identify the brighter features  (i.e., higher LDOS values) in Fig. \ref{Fig14} (top) as the red/dark red lines in Fig.  \ref{Fig05} (bottom). Note, however, that the Majorana mode is much more visible that the ``bullk'' states. This is a consequence of the fact that the Majorana mode is localized near the end of the wire, while the other states have relatively small amplitudes in this region. In addition, in multi-band systems certain states associated with low-energy occupied bands  have large amplitudes at the end of the wire and generate very strong contributions to the LDOS. This is illustrated in the middle panel of Fig. \ref{Fig14}, which shows a two-band system at intermediate coupling having a low-energy spectrum  as in Fig. \ref{Fig08} (b).

%%%%%%%%%%%%%%%%%%%%%%%%%%%%
\begin{figure}[t]
\begin{center}
\includegraphics[width=0.48\textwidth]{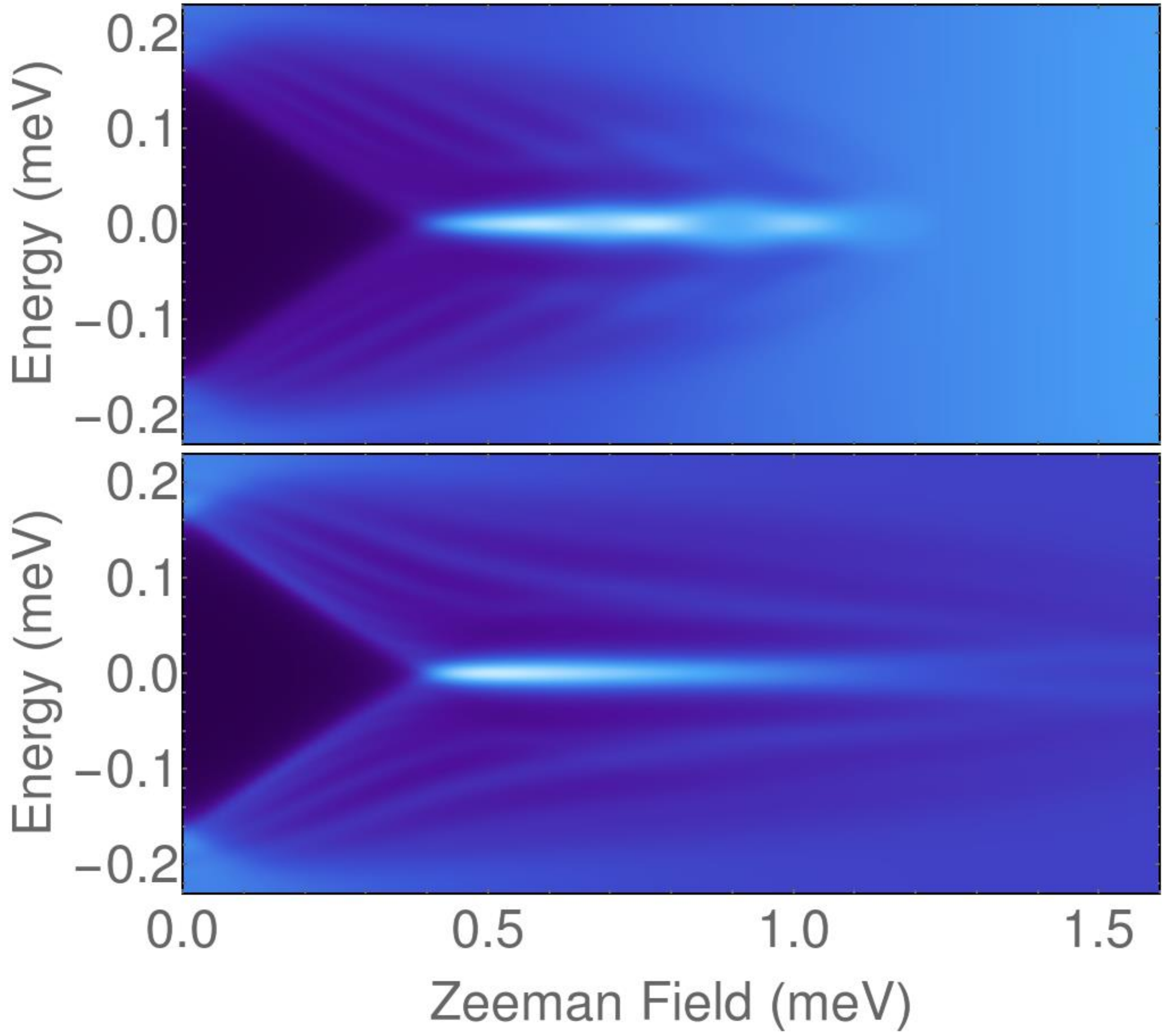}
\vspace{-5mm}
\end{center}
\caption{(Color online) Local density of states for a single band system at intermediate coupling. The model parameters and the color code  are the same as in Fig. \ref{Fig14} (bottom panel). {\em Top}: The bulk SC gap collapses at $\Gamma=1.25$ meV (by contrast, in Fig. \ref{Fig14} and in the bottom panel the gap collapses at $\Gamma=2$ meV). {\em Bottom}: The particle number (rather than the chemical potential) is kept constant.}
\vspace{-2mm}
\label{Fig15}
\end{figure}
%%%%%%%%%%%%%%%%%%%%%%
	
Next, we focus on the second key ingredient that determines the LDOS and consider a ``dirty'' parent superconductor characterized by $\eta(\gamma)$ given by Eq. (\ref{eta}). The dependence of the LDOS on energy and Zeeman field is shown in the bottom panel of Fig. \ref{Fig14}. There are two important features that distinguish the ``clean'' (top) and ``dirty'' (bottom) cases. First, in the ``dirty'' superconductor regime the Majorana mode acquires a significant finite broadening. This introduces a finite energy resolution that limits the observability of the splitting oscillations. Second, and perhaps more importantly, in the presence of a ``dirty'' superconductor with sub-gap states the discrete ``bulk'' spectrum becomes a fuzzy continuum that fills the quasiparticle gap. In other words, if the bulk superconductor has low-energy states in the presence of a finite magnetic field, the Majorana mode is not sharply defined (i.e., it has a finite characteristic width) and is not protected by a quasiparticle gap (i.e., there is a finite density of states at all energies). These features have dramatic consequences regarding the topological properties of the Majorana zero modes. This suggests that  limiting the value $\eta(\Gamma)$ by optimizing the bulk superconductor is a critical requirement for the realization of topological operations with Majorana zero modes.
It is not enough to just have a very clean ballistic SM wire; the parent SC must also be clean, with few subgap states even in the presence of a finite magnetic field, a condition which is hardly satisfied by the SM-SC hybrid structures currently studied experimentally.

The 	final aspect that we address here concerns the visibility of the Majorana splitting oscillations in a hybrid system containing a ``dirty'' superconductor. If the Zeeman field can be varied while maintaining constant chemical potential, the oscillations should be visible in short-enough wires, despite the finite energy resolution introduced by $\eta(\Gamma)$, as evident in the bottom panel of Fig. \ref{Fig14}. Nonetheless, even in this case, the collapse of the bulk SC gap may introduce additional constraints by making the high-field regime inaccessible. This situation is illustrated in the top panel of Fig. \ref{Fig15}, where the Zemman field at which the SC gap collapses has been reduced from  $\Gamma=2$ meV to $\Gamma=1.25$ meV. The splitting oscillations, which are very clear in the ``clean'' system (top panel of Fig. \ref{Fig14}), can barely be resolved. 

In real devices the chemical potential may vary as a function of $\Gamma$, as a result of electrostatic effects. When these effects are strong, it is the particle number (rather than the chemical potential) that is constant as the Zeeman field is varied. This corresponds to a path in the ``phase diagram'' shown in Fig. \ref{Fig12} (bottom) that runs almost parallel to the stripes. This situation is illustrated in the bottom panel of Fig. \ref{Fig15}. Note the absence of splitting oscillations. A faint signature of a single splitting of the zero-bias peak is visible above $\Gamma \approx 1.2$ meV. Of course, if the bulk SC gap collapses at a lower value of the magnetic field (here we have $\Gamma=2$ meV), any signature of the zero-bias peak splitting would completely disappear. 
	
%%%%%%%%%%%%%%%%%%%%%%%%%%%%
\begin{figure}[t]
\begin{center}
\includegraphics[width=0.48\textwidth]{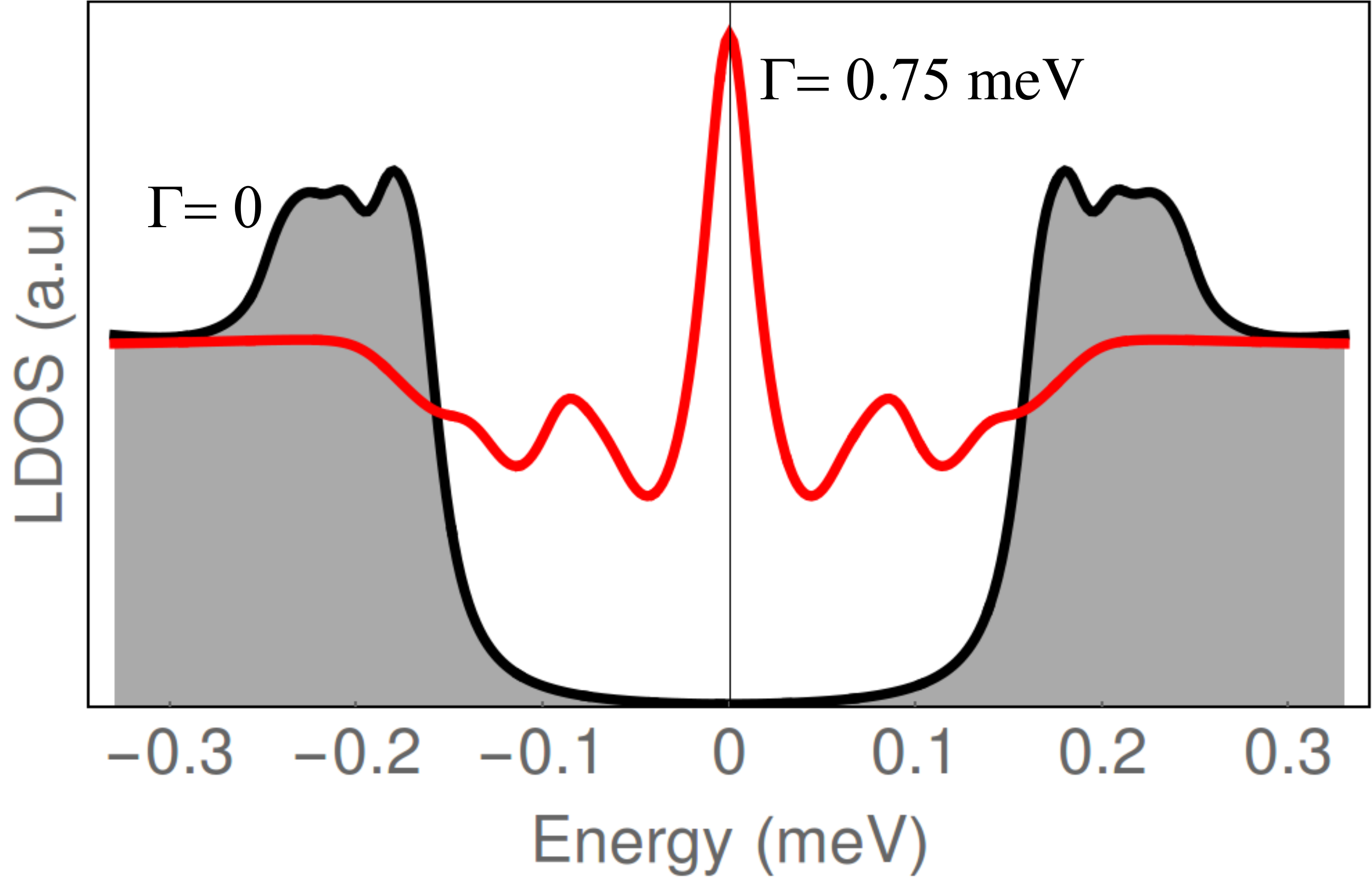}
\vspace{-5mm}
\end{center}
\caption{(Color online) Constant field cuts of the LDOS for a wire with fixed particle number shown in Fig. \ref{Fig15} (bottom panel). A clearly defined ``hard'' gap characterizes the system in the absence of a Zeeman field ($\Gamma=0$). For $\Gamma=0.75$ meV the broad zero-bias Majorana peak cannot be separated from the finite-energy features, which means that the Majorana subspace is not topologically protected by a finite quasiparticle gap.}
\vspace{-2mm}
\label{Fig16}
\end{figure}
%%%%%%%%%%%%%%%%%%%%%%	
	
Finally, we would like to emphasize once more two of the main results of this study: i) The quasiparticle gap that protects the Majorana mode is renormalized as a result of the proximity-coupling to the superconductor. Consequently, in the intermediate and strong coupling regimes the magnitude of the gap is significantly reduced as compared to weak coupling estimates. ii) A ``dirty'' superconductor broadens all the low-energy spectral features of the hybrid system and partially fills in the quasiparticle gap. These properties are illustrated in Fig. \ref{Fig16}, which shows two constant field cuts of the LDOS from Fig. \ref{Fig15} (bottom panel). While in the absence of a Zeeman field the induced gap is clearly defined, at $\Gamma=0.75$ meV one cannot define a quasiparticle gap that separates the broad zero-bias peak from the features associated with finite-energy states. We conclude that obtaining a Majorana mode that is well separated from the finite energy states requires a quasiparticle gap that is much larger that the characteristic energy $\eta(\Gamma)$ for the bulk SC. Consequently, if one uses a low gap superconductor (e.g., Al) one has to work in the intermediate coupling regime (to optimize the induced gap) and to make the bulk SC very clean (to minimize $\eta$). Using a large gap superconductor appears to be a better solution, as it allows one to work in the weak-coupling regime and can tolerate higher levels of sub-gap states (i.e., larger $\eta$). 
We emphasize that the situation depicted in Fig. \ref{Fig16} is rather generic in all SM-SC Majorana systems studied experimentally so far.  In particular, even when a hard gap exists at zero magnetic field, the corresponding gap at a finite field where the Majorana mode may exist (and a zero energy conductance peak is observed), the induced gap is invariably soft.

\section{Conclusions}

We have developed a general theory for proximity-induced topological superconductivity in semiconductor nanowires by fully taking into account the self-energy corrections to the nanowire Green function generated by  the parent superconductor. The theory provides a complete low-energy description of 
 superconductor-semiconductor hybrid structures, which are of much current experimental interest for studying localized  zero energy Majorana modes.  This general theory, which  treats the superconductor and the semiconductor on an equal footing, should be the basis for all future theoretical work in the subject, since recent experiments appear to indicate that the widely used minimal model that only includes Cooper pairs induced by the superconductor into the nanowire and neglects all superconductor-induced renormalization effects, is incapable of describing various key features of the measured data.  
  The theory naturally incorporates the effects of subgap states that may be present in the parent superconductor, particularly in the presence of a finite magnetic field. These effects are shown to have  crucial consequences for the induced  topological superconductivity and the corresponding Majorana modes and provide a natural explanation for the universally observed soft topological gap in semiconductor nanowires.  
  The theory also explains various features in the experimental data as effects arising from the interplay of multi-subband physics,  superconductor-induced renormalization, and superconductor-induced inter-band coupling.  Equally important, the theory shows that the zero-energy nanowire spectrum may manifest complex Majorana splitting oscillations, which cannot be interpreted  on the basis of just the bare semiconductor energy levels, while the topological gap protecting the Majorana modes may be strongly renormalized.
      We establish the quantitative effects of the proximity-induced renormalization on the nanowire topological quantum phase diagram.  Certain features of the phase digram are nonintuitive and much more complex than those predicted by the usual minimal model, which does not incorporate the proximity-induced  renormalization and the inter-band coupling effects. 
       In particular,  inter-subband couplings are non-trivially generated by the parent superconductor, leading to significant modifications in the topological phase diagram. In general, when the nanowire is in the coupled-band regime, no simple equation can describe the quantum transitions between trivial and topological phases.
The theoretical framework provided in our paper should be the starting point in all future efforts to understanding experiments in superconductor-semiconductor hybrid platforms.  

Our work leads immediately to one important conclusion:  If one wants to restrict oneself to the `simple' situation that does not involve `complications' such as soft topological gaps, broad Majorana-related zero bias conductance peaks, and multiple sub-gap structures, which arise from the parent superconductor, then one must take particular care to design hybrid systems characterized by (i) values of the superconductor-semiconductor tunnel coupling much smaller that the gap of the parent superconductor and (ii) the absence of low-energy, sub-gap states in the parent superconductor at relevant values of the magnetic field. Indeed, in the weak-coupling limit the effects of the superconductor-induced renormalization and inter-band coupling are negligible. 
However, if the gap of the parent superconductor is small, this weak-coupling condition comes with the heavy price of the proximity induced gap being very small,  since it will essentially be determined by the weak semiconductor-superconductor coupling.  This would specifically rule out the currently popular InAs-Al core-shell nanowire epitaxial systems, where the tunnel coupling is large (and has to be so to ensure a reasonable value of the induced gap).  If, on the other hand, the parent superconductor has a large gap (e.g., Nb-based superconductors), the size of the induced gap and the energy renormalization effects do not represent major concerns. This is because the  semiconductor-superconductor coupling can be sizable (thus ensuring a large induced gap), yet much smaller than the parent superconducting gap (thus minimizing the renormalization effects). 
  However, even in this case the existence of low-energy sub-gap states in the parent superconductor can easily compromise the properties of the Majorana modes by generating a soft topological gap and creating clusters of hybrid states that appear as broad zero-energy features (e.g., in tunneling measurements). In fact, it is well-known that NbTiN, for example, is an extremely dirty material with substantial disorder-induced subgap states, which makes its use as the parent superconductor very problematic \cite{Driessen2012}.
The actual solution to these problems may require the design of new relatively large-gap parent superconductors with desirable low-energy properties, so that their renormalization of the nanowire spectrum is benign and their subgap states are eliminated.  We urge materials science efforts toward the fabrication of optimal hybrid structures using better parent superconductors in order to avoid the problems of small and soft topological gaps and strongly renormalized low-energy nanowire spectra that plague the current experimental systems.  In particular, the soft topological gap is a very serious problem because a hybrid system with a soft gap is, strictly speaking, gapless or, at best, can be characterized by some small `effective gap'. Since 
the size of the gap itself defines the strength of the topological protection, a small and/or soft topological gap will prevent non-Abelian braiding from being successfu in semiconductor wire-superconductor structures, even if Majorana zero modes are localized at the wire ends.
In our opinion, this is the most important problem preventing further experimental progress in the field.

This work is supported by Microsoft, LPS-MPO-CMTC, and NSF DMR-1414683.				
												
\bibliography{REFERENCES}

\end{document}